\documentclass[aip,cha,amsmath, amssymb,reprint]{revtex4-1}
\usepackage{graphicx}
\usepackage{dcolumn}
\usepackage{bm}
\usepackage{epstopdf}
\usepackage{mathscinet}
\begin{document}
\title{Analytical and numerical study of the non-linear noisy voter model on complex networks}
\author{A. F. Peralta}
\email{afperalta@ifisc.uib-csic.es}
\affiliation{IFISC (Instituto de F{\'\i}sica Interdisciplinar y Sistemas Complejos), Universitat de les Illes Balears-CSIC, 07122-Palma de Mallorca, Spain}
\author{A. Carro}
\affiliation{Institute for New Economic Thinking at the Oxford Martin School, University of Oxford, OX2 6ED, UK}
\affiliation{Mathematical Institute, University of Oxford, OX2 6GG, UK}
\author{M. San Miguel}
\affiliation{IFISC (Instituto de F{\'\i}sica Interdisciplinar y Sistemas Complejos), Universitat de les Illes Balears-CSIC, 07122-Palma de Mallorca, Spain}
\author{R. Toral}
\affiliation{IFISC (Instituto de F{\'\i}sica Interdisciplinar y Sistemas Complejos), Universitat de les Illes Balears-CSIC, 07122-Palma de Mallorca, Spain}
\begin{abstract}
We study the noisy voter model using a specific non-linear dependence of the rates that takes into account collective interaction between individuals. The resulting model is solved exactly under the all-to-all coupling configuration and approximately in some random networks environments. In the all-to-all setup we find that the non-linear interactions induce {\slshape bona fide} phase transitions that, contrary to the linear version of the model, survive in the thermodynamic limit. 
The main effect of the complex network is to shift the transition lines and modify the finite-size dependence, a modification that can be captured with the introduction of an effective system size that decreases with the degree heterogeneity of the network. While a non-trivial finite-size dependence of the moments of the probability distribution is derived from our treatment, mean-field exponents are nevertheless obtained in the thermodynamic limit.
These theoretical predictions are well confirmed by numerical simulations of the stochastic process.
\end{abstract}
\date\today
\maketitle
\begin{quotation}
Imitation models where individuals copy the actions or opinions of others are the basis to understand the transition to consensus and organized behavior in societies. The voter model incorporates the simplest mechanism of blind imitation and has become one of the accepted paradigms in this field. To make the model closer to reality, one needs to take into account the addition of spontaneous changes of state, the characteristics of the network of interactions and the details of the imitation mechanism. All these three ingredients have been considered in the present paper. The resulting non-equilibrium model turns out to offer a rich phenomenology and its phase diagram includes tricritical points, catastrophes, and a non-trivial scaling behavior that can be analyzed using some tools of equilibrium statistical mechanics.
\end{quotation}
\maketitle
\section{Introduction}
The original voter model~\citep{Clifford,Holley} implements in its simplest way the mechanism of random imitation by which a ``voter'' (an individual represented by a variable that can be in any of two possible states) adopts the state of one of its neighbors in a given network of interactions. The voter model displays a stochastic dynamics with absorbing states that correspond to the collective consensus in each of the two possible states and that, from the point of view of Statistical Physics and Critical Phenomena, has very special characteristics~\citep{Hammal}. Some studies~\citep{clopez,Castellano} have examined in detail the robustness or generic behavior of the voter model dynamics by the introduction of nonlinear variations of the random imitation mechanism that, however, respect the existence of the two absorbing states of the stochastic dynamics. A different modification of the voter model is the noisy voter~\citep{Granovsky} model, also called Kirman model~\citep{Kirman}, that has appeared under different names and contexts in the literature~\citep{Moran,Lebowitz,Fichthorn,Considine,Diakonova}, and that adds to the random imitation mechanism spontaneous switches of state, or noise.
It was introduced, in the social sciences literature, as a simple stochastic process that could explain some experimental features observed in ant colonies, when ants have to choose between two symmetrical food sources, as well as some stylized facts ---or non-Gaussian statistical properties--- observed in financial markets, where traders have to decide whether to buy or sell a given stock. In both cases noise is introduced to allow switches between the two collective consensuses avoiding the existence of absorbing states. With its very simple behavioral rules, the noisy voter model is able to capture the emergence of herding and non-Gaussian statistical properties in those two different contexts~\citep{Alfarano1, Alfarano2}. Furthermore, the model presents an order-disorder transition as a function of the relative importance of the spontaneous switching with respect to the copying mechanism, the order parameter being the fraction of voters that share a common value for the state variable. This is known to be a finite-size noise induced transition\citep{Kirman}, with well characterized critical properties in the thermodynamic limit~\citep{Granovsky}, where the transition disappears ---or rather, it occurs in the limit of vanishing spontaneous switching. Some modifications of the system size scaling of the copying mechanism have been proposed to preserve the existence of the order-disorder transition in the thermodynamic limit~\citep{Alfarano1, Alfarano2}. In this paper we address the question of the robustness of the order-disorder transition of the noisy voter model under nonlinear modifications of the random imitation mechanism, emphasizing the system size scaling of its critical properties.

The voter model can be thought of as based on a dyadic, direct interaction between two neighboring voters so that the probability that a voter changes its state is proportional to the fraction of neighboring voters in the opposite state. It is clear that in some situations a more complex mechanism of group interaction can be relevant. This collective interaction introduces a nonlinearity in which the probability to change state is proportional to a power $\alpha$ of the fraction of neighbors in the opposite state. Similar types of nonlinearities, with minor variations in some cases, as the ones introduced here in the noisy voter model have been previously considered in different contexts~\citep{Nowak,Nettle,Strogatz,Castellano,Schweitzer,Vazquez1,Nyczka,Nyczka-2,Min,Jedrzejewski,Moretti,Lambiotte,Molofsky}.
 For example, the degree of nonlinearity, $\alpha$, is assumed in social 
impact theory~\citep{Nowak,Nettle} to be a positive real number measuring the nonlinear effect of local majorities, 
while the same parameter is termed ``volatility'' and is represented by the letter ``$a$" in problems of language competition dynamics~\citep{Strogatz,Vazquez1}. In the interpretation of~\citep{Castellano,Nyczka,Jedrzejewski}, each individual interacts with a set of $\alpha$ of its nearest neighbors, and therefore $\alpha>1$ and it takes integer values. In order to allow for a detailed study of the robustness of the transition of the noisy voter model ---that is, near $\alpha=1$---, we focus here on continuous values of $\alpha$.

We show in this paper that the noisy voter model is structurally unstable with respect to non-linearity so that its order-disorder transition becomes now well defined in the thermodynamic limit for any value of $\alpha>1$. The resulting rich phase diagram, with the presence of tricriticality and catastrophe transitions, extends to the case of continuous $\alpha$ and asymmetry in the transition rates, as some previous results recently presented in the literature~\citep{Nyczka,Jedrzejewski} seem to display. Using analytical techniques and numerical simulations, we carry out a detailed analysis of the dependence with system size of the moments and maxima of the probability distribution of the order parameter. In particular, we focus on the all-to-all interaction and on highly heterogeneous networks of interactions, and we obtain the critical exponents of the finite-size scaling behavior. Furthermore, we prove that the system size scaling of the copying mechanism proposed in~\citep{Alfarano1, Alfarano2} for the noisy voter model is not applicable for nonlinear interactions. In the context of language competition~\citep{Strogatz,Vazquez1}, our study can be understood as an analysis of the robustness of the transition between language coexistence and extinction under noisy perturbations. In this sense, we find that the transition from language extinction to language coexistence at $\alpha=1$ occurs now at a value modified by noise. Interestingly, we also find that there is a tricritical point~\citep{Huang} such that for $\alpha>5$ there is a range of parameters with a new phase of coexistence of three solutions: extinction of each of the two languages or language coexistence in which, in the absence of stochasticity, the initial condition determines the final state.
 
The paper is organized as follows: in Section \ref{model} we define the model, set the notation and introduce the non-linear interaction term. In Section \ref{all-to-all} we solve the model analytically in an all-to-all scenario. This includes the identification of transition lines and the statistical properties of the global state of the system. Section \ref{complex} deals with the characterization of the model in a network structure using the pair-approximation, which is capable of reproducing the main differences with respect to the all-to-all result. We end with a summary and conclusions in Section \ref{conclusions}

\section{Model}\label{model}
We consider a population of $N$ individuals located in the nodes of a given (undirected) network of interactions. Each node holds a time-dependent binary variable $n_{i}=0,1$, defining the state of individual $i=1,..., N$. The meaning of this binary variable does not concern us in this paper, but typical interpretations include the optimistic/pessimistic state of a stock market broker~\citep{Kirman}, the language A/B used by a speaker~\citep{Strogatz,Vazquez1}, or the direction of the velocity right/left in a one-dimensional model of active particles~\citep{Escaff}. The network of interactions is mapped onto the usual (symmetric) adjacency matrix of coefficients $A_{ij}=1$ if nodes $i$ and $j$ are connected and $A_{ij}=0$ otherwise. The degree of node $i$ is its total number of connected nodes (also known as neighbors) $k_i=\sum_{j} A_{ij}$. An important characteristic of the network is its degree distribution, $P_{k} \equiv N_{k}/N$, where $N_{k}$ is the number of nodes with degree $k$. The $m$-moment of the degree distribution is defined as $\mu_{m}=\sum_{k} P_{k} k^m$, with short notation $\mu \equiv \mu_1$.

The state of a node can change over time following the combination of a purely random, or noisy, effect and a herding, or copying, mechanism.  They are both stochastic processes defined as:\\
1.- The purely random effect takes into account that individuals can change their state independently of the state of others, with a rate $a_0$ or $a_1$, depending on whether the node holds state $0$ or $1$, respectively. This process reflects an idiosyncratic or autonomous behavioral rule.\\
2.- The herding mechanism considers that individuals can be persuaded by their neighbors to change their state. Hence, individual $i$ in state $n_i=0$ can change the state to $n_i=1$ with a rate $h\, f \left[ \sigma_{1}(i) \right]$. Here $\sigma_{1}(i)$ is the fraction of neighbors of $i$ holding the opposite state $1$, $h$ is a parameter measuring the herding intensity, and $f$ is a monotonically increasing function that fulfills $f[0]=0$ and $f[1]=1$. Similarly, an individual in the state $n_i=1$ can change to state $n_i=0$ with a rate $h \, f \left[ \sigma_{0}(i) \right]$, where $\sigma_{0}(i)=1-\sigma_1(i)$.

The total individual transition rates $\pi_{i}^{\pm}$ result from the sum of these two complementary processes
\begin{eqnarray}
\label{ind_rates+}
\pi_{i}^{+} &\equiv& \text{rate}(n_{i}=0 \rightarrow n_{i}=1) = a_{0} + h f[\sigma_{1}(i)],\\
\label{ind_rates-}
\pi_{i}^{-} &\equiv& \text{rate}(n_{i}=1 \rightarrow n_{i}=0) = a_{1} + h f[\sigma_{0}(i)].
\end{eqnarray}
In terms of the of the adjacency matrix, $\sigma_{0/1}(i)$ can be written as:
\begin{equation}
\label{sigma}
\sigma_{1}(i) = \frac{1}{k_{i}} \sum_{j=1}^{N} A_{ij} n_{j}, \hspace{10pt} \sigma_{0}(i) = 1-\sigma_{1}(i).
\end{equation}

The function $f[\sigma]$ depicts the nature of the copying mechanism. For example, the traditional voter model uses a linear dependence $f[\sigma]=\sigma$ which corresponds to the process in which a node copies the state of a randomly selected neighbor. We will focus, however, on a more general way to model the interactions, considering a non-linear function $f[\sigma]=\sigma^{\alpha}$. For $\alpha>1$ individuals are more resistant to follow the opinion of the neighbors holding the opposite state, while the contrary is true for $\alpha<1$, it is easier to copy the opposite state of a neighbor. Hence, a value of $\alpha>1$ (probability of imitation below random) can be representative of a situation of aversion to change, where a larger fraction of neighbors in the opposite state is needed to switch state compared to the purely random imitation of the voter model. In fact, values of $\alpha>1$ have been fitted in some problems of language competition\cite{Strogatz}. On the other hand, values of $\alpha<1$, corresponding to a probability of imitation above random or a situation of preference for change, have been considered in social impact theory~\citep{Nowak}. The cases $\alpha=2,3...$ are equivalent to a process in which an individual changes state if and only if after checking repeatedly the state of $\alpha$ randomly selected neighbors (and allowing for repetitions in the selection), all of them happen to be in the opposite state to the one held by the individual~\citep{Castellano}. 

The global state of the system can be characterized by the total number of nodes in state $1$, $n=\sum_{i=1}^{N} n_{i}$, taking integer values $n \in [0, N]$ or, by the intensive variable (also known as ``magnetization'') $m = 2 n/N-1$ that takes values in the range $m \in [-1,+1]$. The values $m=-1$ and $m = +1$ correspond to a situation where all the nodes hold the same state, respectively $n_i=0$ and $n_i=1$, while $m=0$ is the perfectly balanced case, where half of the nodes are in state $n_i=0$ and the other half in $n_i=1$. One of the particularities of the noisy version of the voter model is that for $a_{0} > 0$ the system, even after long enough times, does not get stuck at $n=0$, nor at $n=N$ for $a_1>0$, the two absorbing states of the noiseless voter model.

The aim of this paper is to study the stationary statistical properties of the magnetization, namely its moments $\langle m^k\rangle$ and the location of the maxima of its probability distribution as a function of the parameters $a_0, a_1, h, \alpha$, as well as the population size $N$ and the structure of the network of interactions, with special attention to the role of the non-linear parameter $\alpha$. In order to do that, in the next sections, we will develop analytical results that will be compared with those coming from numerical simulations of the stochastic process using well-known techniques\cite{f1}.

\section{All-to-all solution}\label{all-to-all}
In the all-to-all scenario, all the nodes are equivalent, and the only relevant variable is $n$. In this sense, we do not consider the individual rates Eqs.(\ref{ind_rates+},\ref{ind_rates-}) but the global rates $\pi^{\pm}(n)\equiv \text{rate}(n \rightarrow n \pm 1)$ at which a change of state in any of the nodes takes place. The individual fraction $\sigma_{1}(i)$ is replaced by the global fraction $n/N$ (while $\sigma_{0}(i)$ by $(N-n)/N$) and thus the global rates read:
\begin{eqnarray}
\label{glob_rates+}
\pi^{+}(n) &=& (N-n) \left( a_{0} + h \left( \frac{n}{N} \right)^{\alpha} \right),\\
\label{glob_rates-}
\pi^{-}(n) &=& n \left( a_{1} + h \left(\frac{N-n}{N} \right)^{\alpha} \right).
\end{eqnarray}

The approach is exact only when the network is fully connected, namely when each node is connected to all the other nodes. The transition rates Eqs.(\ref{glob_rates+},\ref{glob_rates-}) define a Markovian stochastic process, which is of the one-step type. The process can be fully described by means of the probability $P(n, t)$ to have $n$ individuals in state $1$ at time $t$, which obeys the master equation~\citep{VanKampen:2007}:
\begin{equation}
\label{Master_equation}
\frac{\partial P(n,t)}{\partial t} = \left(E^{+}-1 \right) \left[ \pi^{-}(n) P \right] + \left(E^{-}-1 \right) \left[ \pi^{+}(n) P \right],
\end{equation}
where we have introduced the step operators $E^{\pm}$, defined to act on an arbitrary function $g(n)$ as $E^{\pm}[g(n)]=g(n \pm 1)$. One can derive a continuous version of this equation, known as Fokker-Planck equation, through a systematic expansion in $N$. Rewriting Eq.(\ref{Master_equation}) in terms of the magnetization $m$ and expanding it in power series of $\Delta m = \pm 2/N$ to second order, one finds
\begin{equation}
\label{Fokker-Planck_equation}
\frac{\partial P(m,t)}{\partial t} = - \frac{\partial}{\partial m} \left[ F(m) P \right]+\frac{1}{N} \frac{\partial^2}{\partial m^2} \left[ D(m) P \right],
\end{equation}
where $P(m,t)$ is now a probability density function (pdf), related to the discrete probability function as $P(n,t)=P(m,t) \cdot \vert d m / dn \vert=P(m, t) \cdot 2/N$. The functions $F(m) = \left[ \pi^{+} - \pi^{-} \right] 2/N$ and $D(m)= \left[ \pi^{+} + \pi^{-} \right] 2/N$ are usually called drift and diffusion respectively and, for this model, are given by
\begin{align}
\label{drift}
F(m) &= a_0 - a_1 - (a_0+a_1) m \\
 &+ 2^{-\alpha} h (1-m^2) \left( (1+m)^{\alpha-1} - (1-m)^{\alpha-1} \right),\notag\\
\label{diffusion}
D(m) &= a_0 + a_1 - (a_0-a_1) m \\
 &+ 2^{-\alpha} h (1-m^2) \left( (1+m)^{\alpha-1} + (1-m)^{\alpha-1} \right).\notag
\end{align}
It is also possible to describe the evolution of the system in terms of a Langevin stochastic differential equation for the trajectories $m(t)$, which, within the It\^{o} convention, reads
\begin{equation}
\label{Langevin_equation}
\frac{d m(t)}{d t}= F(m) + N^{-1/2} \sqrt{2 D(m)} \cdot \xi(t),
\end{equation}
where $\xi(t)$ is a Gaussian white noise with zero mean $\langle \xi(t) \rangle = 0$ and correlations $\langle \xi(t) \xi(t') \rangle = \delta(t-t')$.
\subsection{Modes and transitions}
In general, it is difficult to find an analytical solution of the master equation (\ref{Master_equation}) or the Fokker-Planck equation (\ref{Fokker-Planck_equation}). There is, however, a general solution $P_{\rm st}(m)$ of the latter\cite{f2} in the stationary state $(\partial / \partial t) P(m, t)=0$ that can be written in the exponential form\begin{eqnarray}
\label{Fokker-Planck_solution}
P_\text{st}(m) &=& {\cal Z}^{-1} \cdot \exp \left[ -N V(m) \right], \\
\label{Potential}
V(m) &=& \int^{m} \frac{-F(z)+D'(z)/N}{D(z)} dz,
\end{eqnarray}
where $V(m)$ is called the ``potential function'', and $\cal Z$ is a normalization constant.

Although the solution Eqs.(\ref{Fokker-Planck_solution},\ref{Potential}) is a complicated expression given the drift and diffusion functions Eq.(\ref{drift},\ref{diffusion}), it is possible to portray the shape of the stationary distribution $P_\text{st}(m)$. In particular, in this subsection we will focus on characterizing the modes, or values of $m_{*}$ for which $P_\text{st}(m_{*})$ is a maximum, and the transitions where modes appear or disappear as a function of the parameters of the model.

The condition of local maximum for the mode is $P_\text{st}'(m_{*})=0$ (condition of local extreme) and $P_\text{st}''(m_{*})<0$, but note that modes can also be located at the boundary values $m_{*}=\pm1$. In the symmetric case $a_0=a_1$ there is a trivial extreme $m_{*}=0$ which corresponds to the perfectly balanced case. Its dynamical stability under perturbations comes determined by the second derivative of $P_\text{st}(m)$ or, alternatively, of $V(m)$
\begin{equation}
\label{stability_zero}
V''(m_{*}=0)=\frac{2^{\alpha}}{2^{\alpha} \varepsilon +1} ( \varepsilon - \varepsilon_c(N) ),
\end{equation}
where we define the noise-herding intensity ratio parameter $\varepsilon \equiv (a_0+a_1)/2h$, and $\varepsilon_c(N)$ reads
\begin{eqnarray}
\label{critical_temperatureN}
\varepsilon_c (N) &=& 2^{-\alpha} \left( \alpha-1+\frac{\alpha (3-\alpha)}{N} \right),\\
\varepsilon_c &\equiv& \varepsilon_c (\infty) =2^{-\alpha} \left( \alpha-1 \right).\label{critical_temperature}
\end{eqnarray}
Then, for $\varepsilon < \varepsilon_c(N)$, $m_{*}=0$ corresponds to a minimum of the probability distribution, while for $\varepsilon > \varepsilon_c(N)$ to a maximum. Note that in the thermodynamic limit $N\to\infty$ the transition between maximum and minimum occurs at a finite positive value of $\varepsilon$ for $\alpha > 1$, while for $\alpha < 1$ there is no transition since $\varepsilon_c(\infty) < 0$ and $m_{*}=0$ is always a stable solution. For the particular value $\alpha=1$, the linear model, the transition is only a finite-size effect since $\varepsilon_c(N)=1/N$ and, in the thermodynamic limit, the maximum is located at $m_{*}=0$ for all values of $\varepsilon>0$. 

Consequently, only in the non-linear regime $\alpha>1$, the parameters $a_{0}$, $a_1$ and $h$ take values of the same order near the transition, and we can safely disregard the term $D'(z)/N$ in front of $F(z)$ in the numerator of Eq.(\ref{Potential}), in the limit of large $N$. In this limit, that we assume in the remaining of this subsection, the extrema of the distribution fulfill the condition $F(m_{*})=0$ while the stability comes determined by the sign of the derivative $F'(m_{*})$. Since it is not possible to find a closed solution of this equation, we will expand the drift function in power-series to $O(m^7)$
\begin{align}
\label{drift_expand}
F(m)&= h \Delta\varepsilon + 2 h \left( \varepsilon_{c}- \varepsilon \right) m+ \frac{h}{3} 2^{-\alpha} (\alpha-5) (\alpha-1) \alpha m^3 \notag\\
&+ \frac{h}{15} 2^{-2-\alpha} (\alpha-9) (\alpha-3) (\alpha-2) (\alpha-1) \alpha m^5\notag\\ &+O(m^7),
\end{align}
where we define the parameter $\Delta\varepsilon = (a_0-a_1)/h$ as a measure of the asymmetry in the rates for switching states $0\to1$ and $1\to 0$. This expression coincides with the derivative of the potential function found by Vazquez et al.~\citep{Vazquez1} in their study of the noiseless version ($\varepsilon= \Delta \varepsilon =0$) of the nonlinear model in the unbiased case\cite{f3}.
In principle, the solution (\ref{drift_expand}) using the power-series expansion will be valid only for $m_{*} \approx 0$, but it is enough to identify additional extrema and transitions. We now find the extrema and their stability distinguishing between the symmetric $a_0=a_1$ and asymmetric $a_0\ne a_1$ cases.

\begin{figure}[h]
\centering
\includegraphics[width=0.5\textwidth]{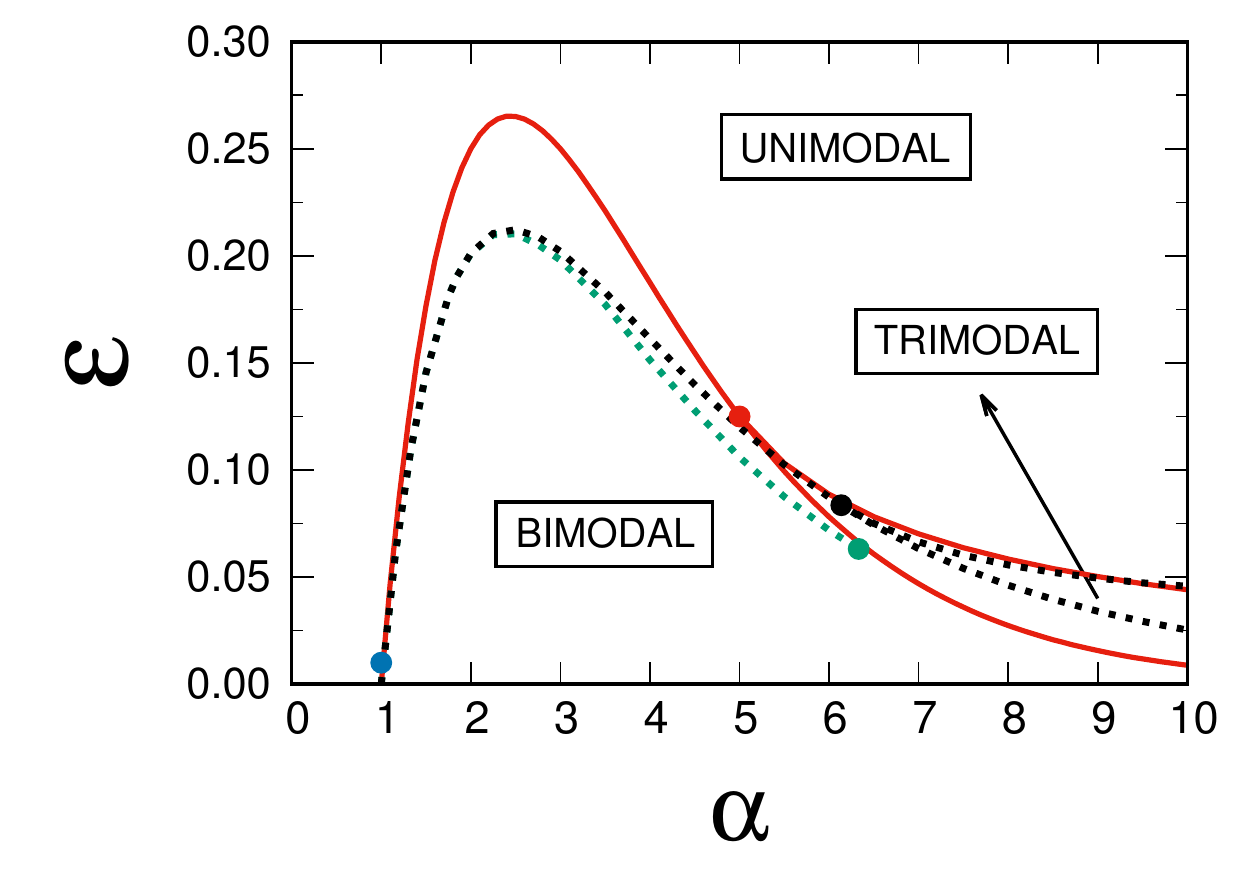}
\caption{Phase diagram of the symmetric case $\Delta \varepsilon = 0$. The dot (blue) at $\alpha=1$ is the finite-size transition point $\varepsilon_c(N)$, Eq.(\ref{critical_temperatureN}) for $N=100$. Solid line corresponds to the all-to-all setting, while the dotted line to a $15$-regular network (black) and a scale free network $P_{5 \leq k \leq 966} \sim k^{-2.34}$ with $\mu \approx 15$ (green), see Section \ref{complex}. The trimodal region is delimited by the transition lines $\varepsilon_{c}$ and $\varepsilon_{t}$, which for the all-to-all network correspond to expressions Eqs.(\ref{critical_temperatureN},\ref{critical_temperature_2}), with a tricritical  point (red) at $\alpha=5,\,\varepsilon=1/8$. The tricritical point for the $15$-regular network (black) is at $\alpha=6.14,\,\varepsilon=0.084$ and for the scale free network (green) at $\alpha=6.33,\,\varepsilon=0.063$ (the trimodal region of the scale free network is removed for clarity in the figure).}
\label{fig:modes1}
\end{figure}

\begin{figure}[h]
\centering
\includegraphics[width=0.45\textwidth]{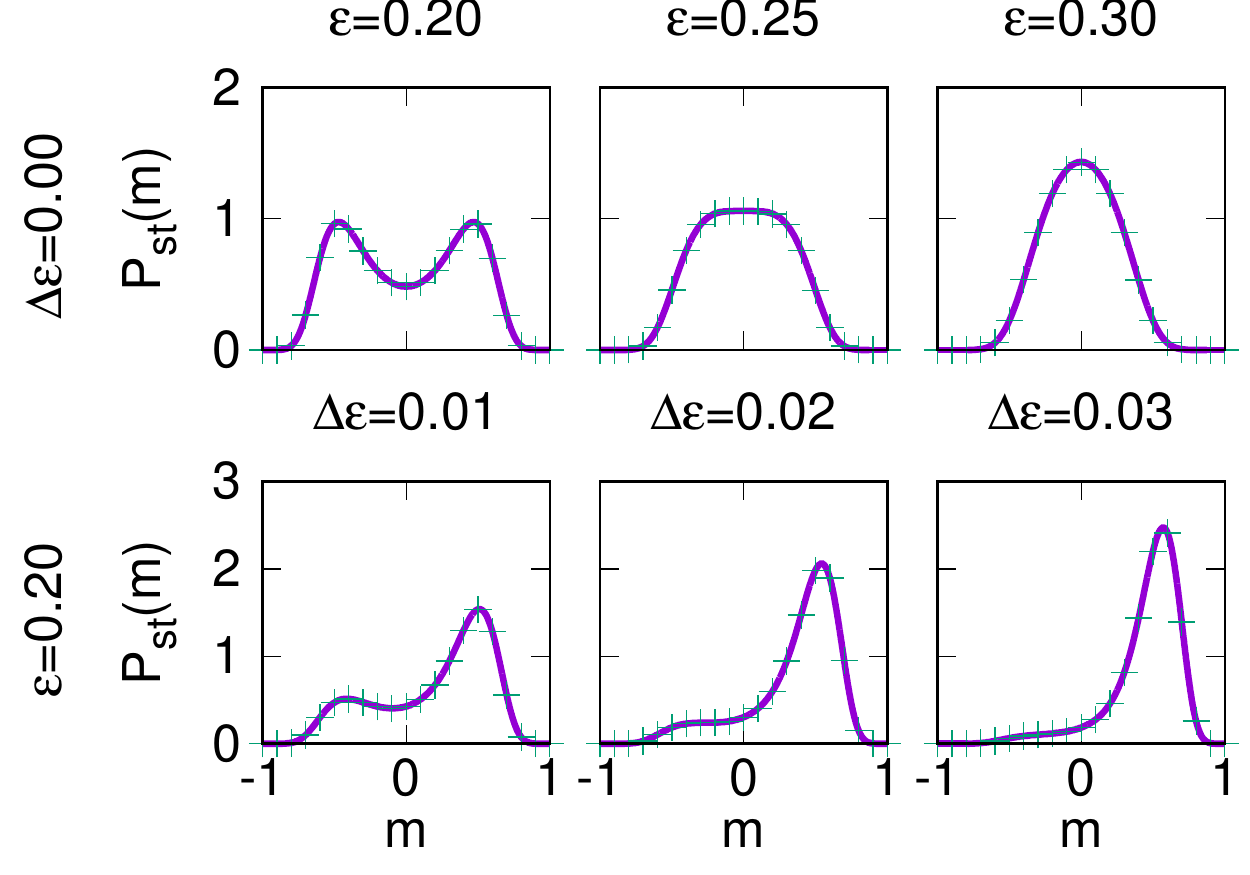}
\caption{Probability distribution $P_{\text{st}}(m)$, for different values of the parameters with fixed $\alpha=2$. In the symmetric case $\Delta\varepsilon=0$ (top panels) the transition point is at $\varepsilon_{c}=0.25$ where the distribution switches from symmetric bimodal to unimodal. In the bottom panels we identify the transition from an asymmetric bimodal (with two maxima, one of them absolute and the other local) to a unimodal distribution occurring at a value of the asymmetry parameter $\Delta\varepsilon_{a} =0.017$ for $\varepsilon=0.2$ (see the corresponding points in Fig.\ref{fig:modes2}). Dots correspond to numerical simulations of the process defined by the rates Eqs.(\ref{glob_rates+},\ref{glob_rates-}) while lines are the function Eq.(\ref{Fokker-Planck_solution}).}
\label{fig:modes3}
\end{figure}

\begin{figure}[h]
\centering
\includegraphics[width=0.45\textwidth]{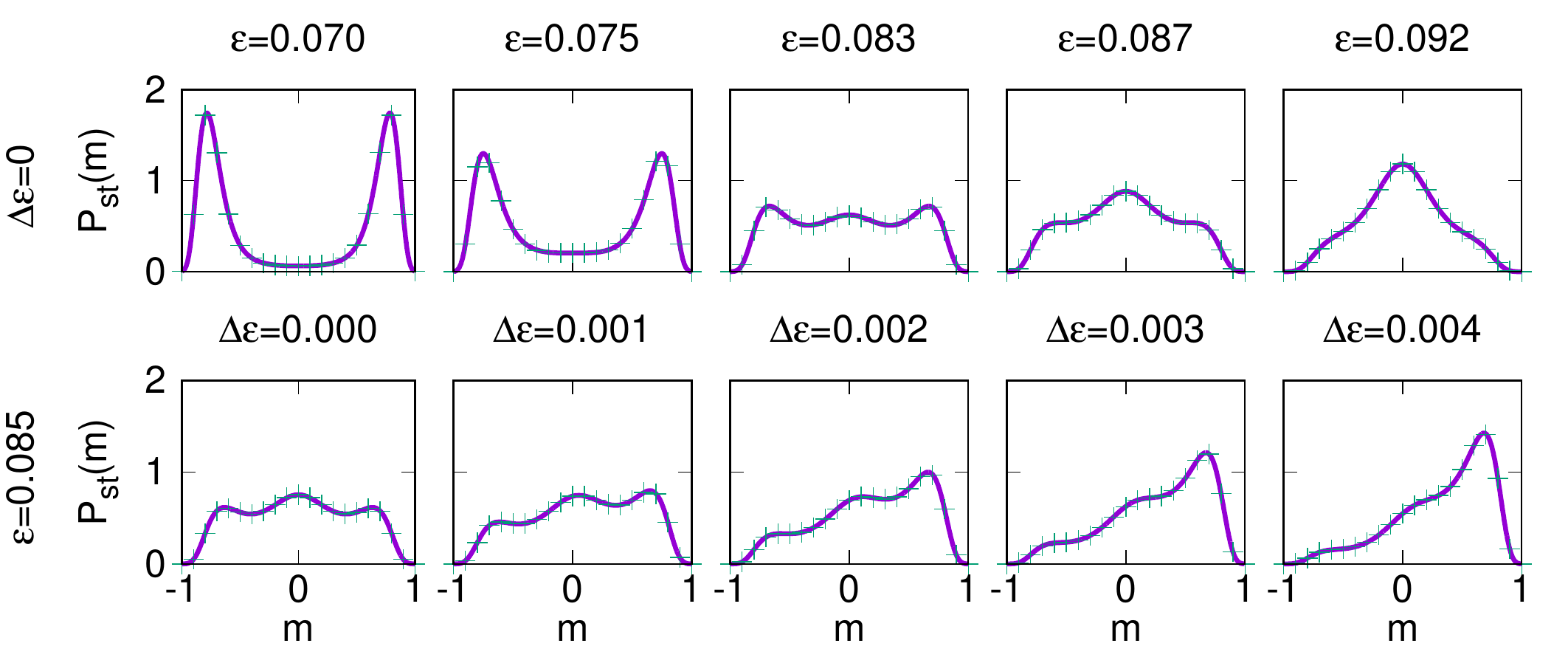}
\caption{Probability distribution $P_{\text{st}}(m)$, for different values of the parameters with fixed $\alpha=6$. In the symmetric case $\Delta\varepsilon=0$ (top panels) the transition points are $\varepsilon_{c}=0.075$ (from bimodal to trimodal), and $\varepsilon_{t}=0.086$ (from trimodal to unimodal). In the bottom panels we fix $\varepsilon=0.085$ and identify the transition from a trimodal to a bimodal distribution occurring at a value of the asymmetry parameter$\Delta\varepsilon_{a,1} = 0.00082... $, and the transition from bimodal to unimodal occurring at $\Delta\varepsilon_{a,2}= 0.00280...$ (see the corresponding points in Fig.\ref{fig:modes2}). Dots correspond to numerical simulations of the process defined by the rates Eqs.(\ref{glob_rates+},\ref{glob_rates-}) while lines are the function Eq.(\ref{Fokker-Planck_solution}).}
\label{fig:modes4}
\end{figure}

\subsubsection{Symmetric case} Using the expansion Eq.(\ref{drift_expand}) there are five (maybe complex) solutions of $F(m_{*})=0$. When $a_0=a_1$, they are as follows: the trivial one $m_{*}=0$ and other four roots $\pm m_{*}^{+}$, $\pm m_{*}^{-}$ obtained from:
\begin{align}
\label{extrema}
 &(m_{*}^\pm)^2 \approx \frac{10 (\alpha -5) }{(\alpha-2) (\alpha-3) (\alpha-9)} \cdot \\
 &\left[ \pm \sqrt{1-\frac{6}{5} \frac{(\alpha-2)(\alpha-3)(9-\alpha)}{\alpha (\alpha-5)^2} \left( \frac{\varepsilon-\varepsilon_{c}}{\varepsilon_{c}} \right)} - 1 \right].\notag
\end{align}
The acceptable maxima of the probability distribution $P_\text{st}(m)$ must correspond to values of $m_{*}$ which are real and inside the interval $[-1,1]$. The four roots given by Eq.(\ref{extrema}) are real or imaginary depending on the values of the parameters $\varepsilon$ and $\alpha$:

\begin{enumerate}
\item In the range $1<\alpha<5$, for $\varepsilon < \varepsilon_{c}$ the pair of solutions $\pm m_{*}^+$ are real and correspond to probability maxima, while for $\varepsilon > \varepsilon_{c}$ all four roots are imaginary. Then, in this range, the line $\varepsilon_{c}$ divides unimodal (a single maximum of the probability distribution) from bimodal (two real maxima) regimes, see Fig.\ref{fig:modes1} and top panels of Fig.\ref{fig:modes3}.
\item In the range $\alpha>5$, for $\varepsilon < \varepsilon_{c}$ the pair of solutions $\pm m_{*}^+$ are real and correspond to probability maxima; for $\varepsilon_{c} < \varepsilon < \varepsilon_{t}$ all four roots are real, $\pm m_{*}^-$ correspond to probability maxima, and $\pm m_{*}^+$ to minima; while for $\varepsilon > \varepsilon_{t}$ all four are imaginary, where $\varepsilon_{t}$ reads
\begin{equation}
\label{critical_temperature_2}
\varepsilon_{t} \approx \varepsilon_{c} \cdot \left(1+\frac{5 \alpha (\alpha-5)^2}{6 (\alpha-2)(\alpha-3)(9-\alpha)} \right).
\end{equation}
Consequently, for $\alpha>5$, the line $\varepsilon_{c}$ divides bimodal from trimodal regimes, while $\varepsilon_{t}$ divides trimodal from unimodal regimes, see Fig.\ref{fig:modes1} and the top panels of Fig.\ref{fig:modes4}. The two lines meet at the tricritical point $\alpha=5$ and $\varepsilon=2^{-3}=0.125$.
\end{enumerate}
Note that the expressions for the roots Eq.(\ref{extrema}) and the transition line Eq.(\ref{critical_temperature_2}) are based on the expansion Eq.(\ref{drift_expand}) and consequently, they are an approximation, hence the $\approx$ symbol used in those formulas. In fact, we have found that they remain accurate only for $\alpha < 7$. Despite that, the classification of maxima/minima is completely general and additional extrema or transitions are not observed in a numerical analysis using the exact expression for the potential function Eq.(\ref{Fokker-Planck_solution}) and Eqs.(\ref{drift},\ref{diffusion}).

It is also important to mention that for $1 < \alpha < 5$, the line $\varepsilon = \varepsilon_{c}$ corresponds to a classical (Landau) second order phase transition from the ``ordered'' phase $m_{*}\ne 0$ to the ``disordered'' one\cite{f4} $m_{*}=0$, with scaling $\vert m_{*}\vert \sim (\varepsilon_c - \varepsilon )^{\beta}$ for $\varepsilon \le \varepsilon_{c}$ with $\beta=1/2$ and $m_{*}=0$ for $\varepsilon \ge \varepsilon_{c}$. This line is delimited by two special degenerate points, $\alpha=1$ and $\alpha=5$:
\begin{itemize}
\item For $\alpha=1$, all the terms of the expansion Eq.(\ref{drift_expand}) vanish and $P_\text{st}(m)$ is completely flat at the transition. It is only at $\varepsilon=\varepsilon_c=0$ (noiseless voter model) that the probability distribution is a sum of two delta-functions at $m=\pm1$, leading to $m_*=\pm1$ if $\varepsilon=0$ or $m_*=0$ if $\varepsilon>0$. This discontinuity survives in the finite $N$ limit, where the result is $m_*=\pm1$ if $\varepsilon<\varepsilon_c(N)$ or $m_*=0$ if $\varepsilon>\varepsilon_c(N)$. Formally, we still have a scaling of the form $\vert m_{*}\vert \sim (\varepsilon_{c}(N)-\varepsilon)^{\beta}$, but now $\beta=0$.
\item For $\alpha=5$, the first three terms of the expansion vanish and $P_\text{st}(m)$ is flatter than for the classical second order transition, with a scaling $ \vert m_{*}\vert \sim(\varepsilon_c - \varepsilon)^{\beta}$, $\beta=1/4$.
\end{itemize}

The condition that isolates the classical second order line from the degenerate points is that we can disregard the $O(m^5)$ term of Eq.(\ref{drift_expand}) in front of the $O(m^3)$, which leads to
\begin{equation}
\label{classical_second}
\left\vert \varepsilon - \varepsilon_c \right\vert \ll \varepsilon_c \cdot \left\vert \frac{\alpha (\alpha-5)^2}{(\alpha-2)(\alpha-3)(9-\alpha)} \right\vert.
\end{equation}

For $\alpha>5$ an increase of $\varepsilon$ leads first to a transition from a bimodal to a trimodal distribution at $\varepsilon=\varepsilon_c$, implying a discontinuity of the location of the absolute maximum from $\vert m_{*}\vert>0$ to $m_{*}=0$. This corresponds to a first-order transition. The local maxima at $\vert m_{*}\vert>0$ remain up to $\varepsilon=\varepsilon_t$ where they disappear and the distribution is unimodal. The absolute maximum of the distribution is at $\vert m_{*}\vert>0$ for $\varepsilon<\varepsilon_M$ and at $m_{*}=0$ for $\varepsilon>\varepsilon_M$, being $\varepsilon_M$ the Maxwell point where the potential takes the same value at the maxima, $|m_{*}|>0$ and $m_{*}=0$.

\begin{figure}[h]
\centering
\includegraphics[width=0.22\textwidth]{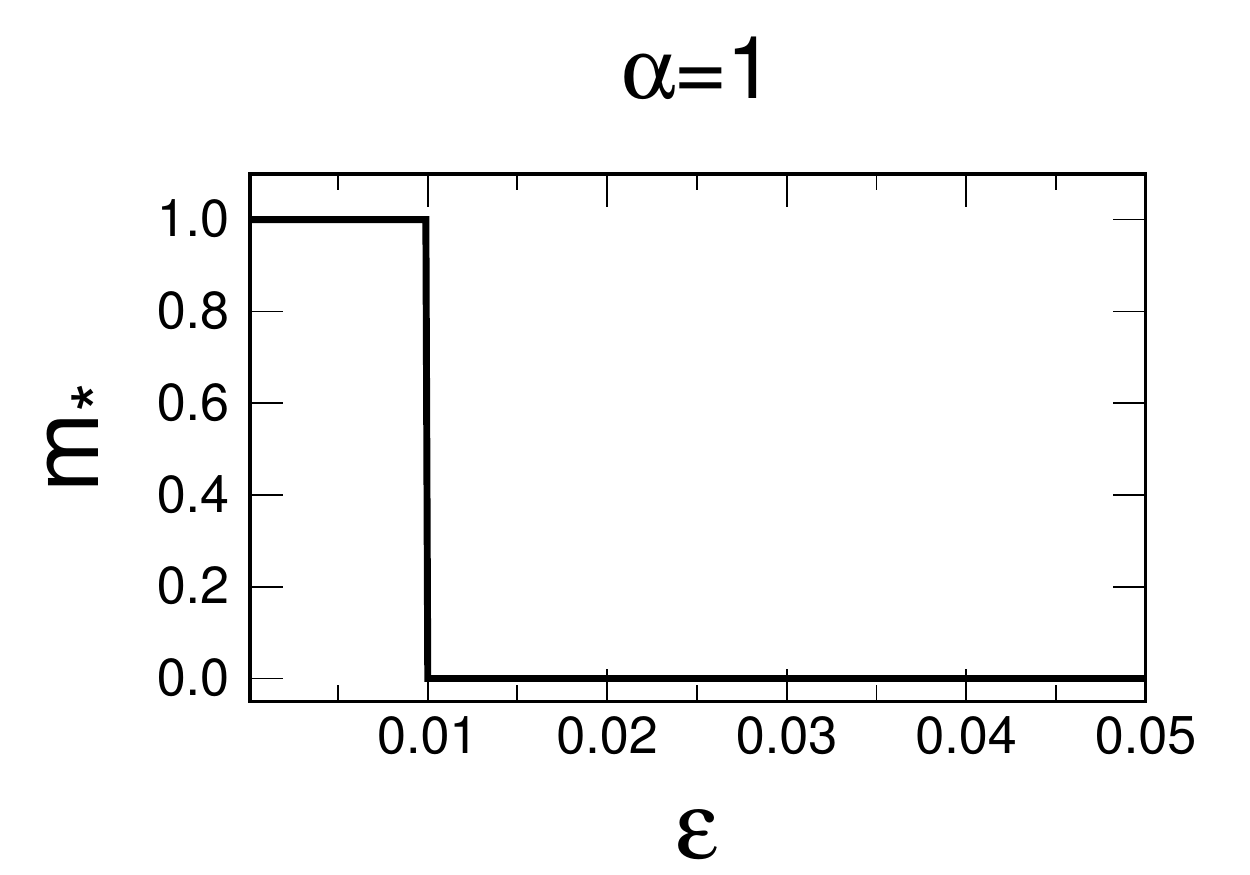}
\includegraphics[width=0.22\textwidth]{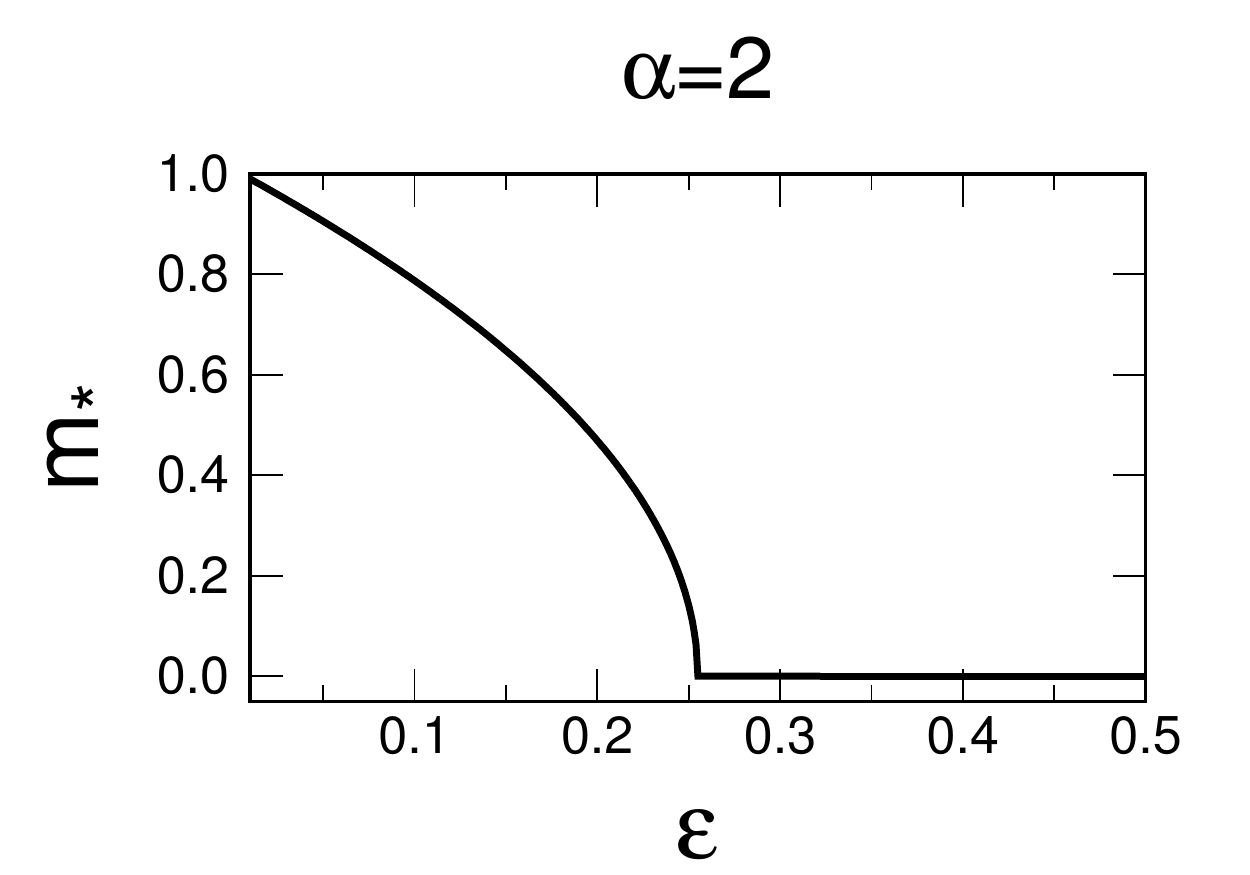}
\includegraphics[width=0.22\textwidth]{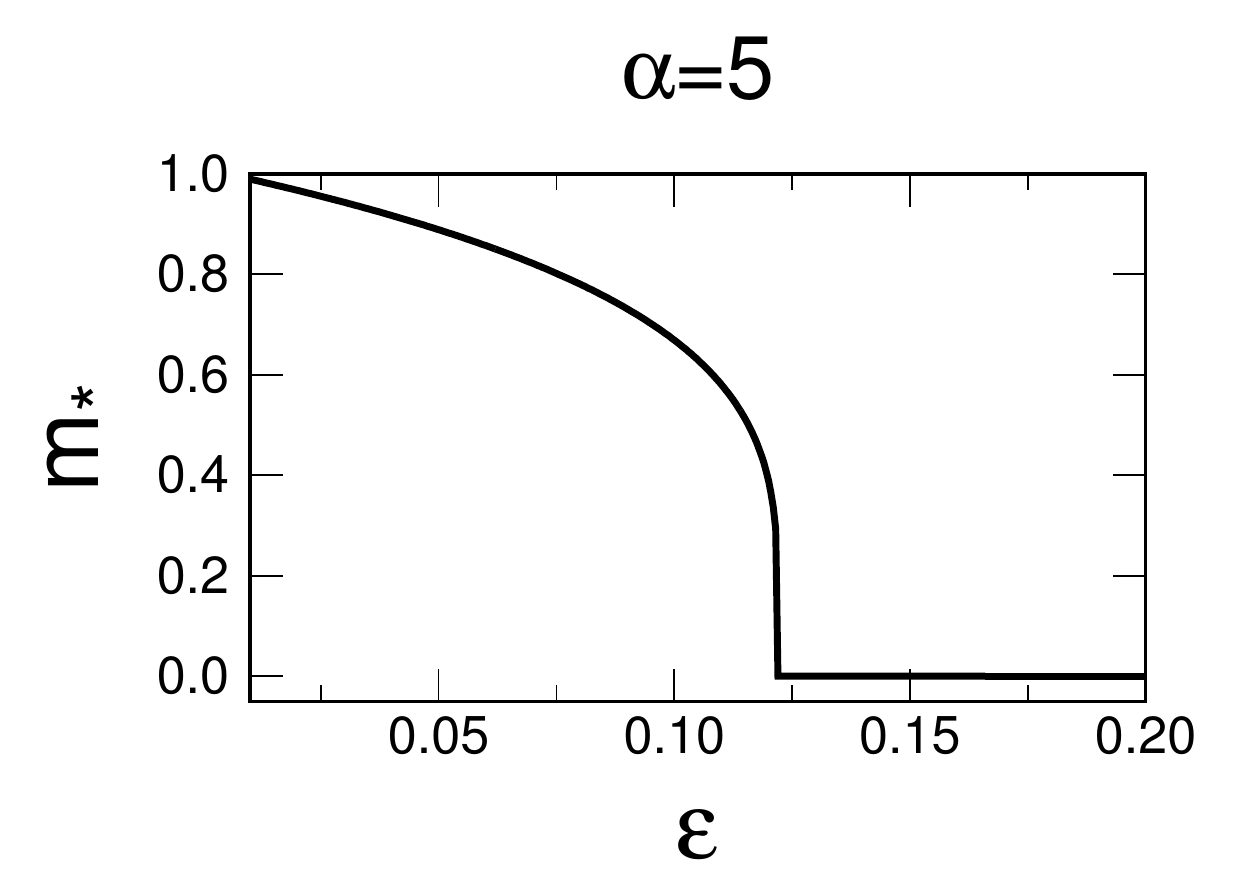}
\includegraphics[width=0.22\textwidth]{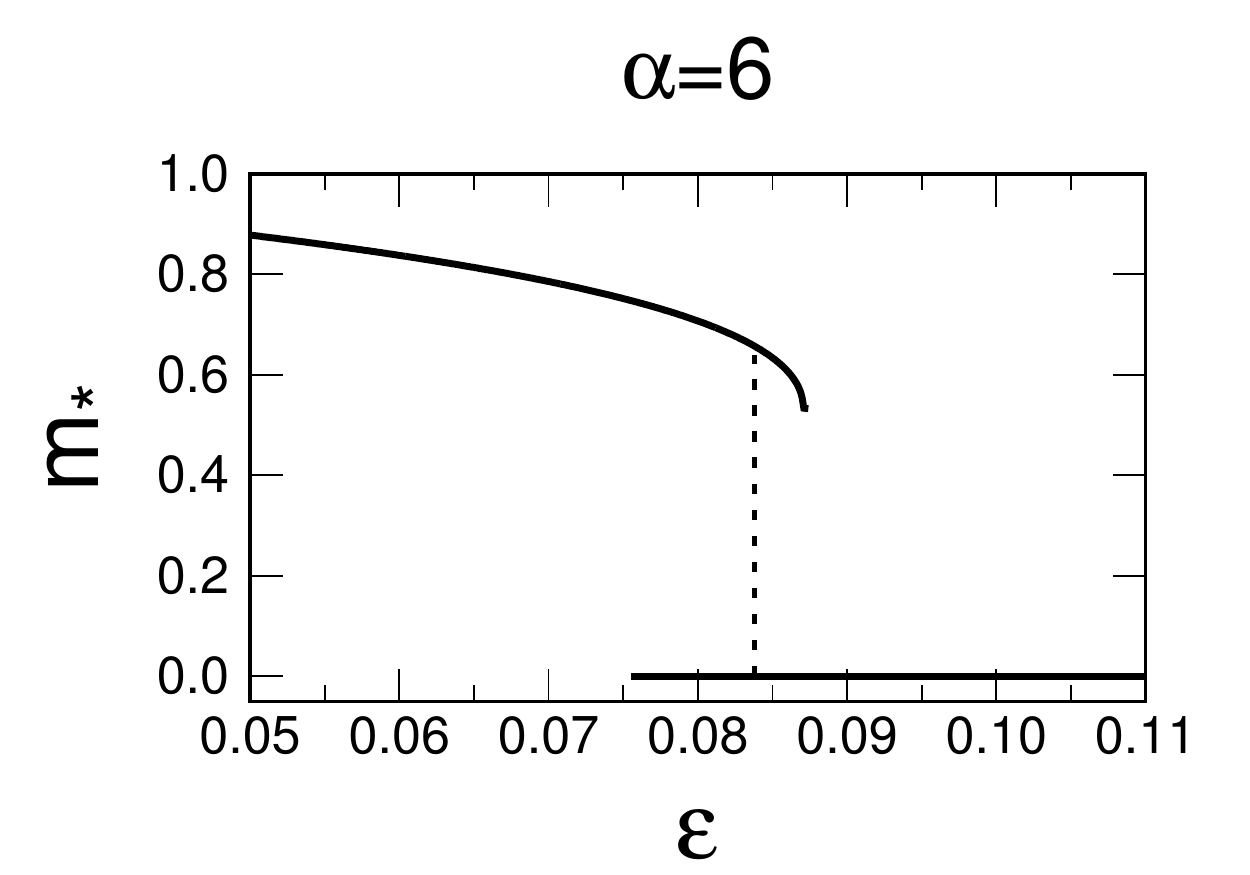}
\caption{Summary of the equation of state for different values of $\alpha$ in the symmetric case $\Delta\varepsilon=0$. The lines come from the determination of the maxima $m_{*}$ of the exact potential Eq.(\ref{Potential}) for $N=100$. The vertical dashed line for $\alpha=6$ is $\varepsilon_M=0.0838$, the Maxwell point where the potential takes the same value at the two maxima, $m_{*}>0$ and $m_{*}=0$.}
\label{fig:phase}
\end{figure}

A summary of the equation of state $m(\varepsilon)$ for different values of $\alpha$ in the symmetric case is presented in Fig.\ref{fig:phase}.

\begin{figure*}[t]
\centering
\includegraphics[width=0.45\textwidth]{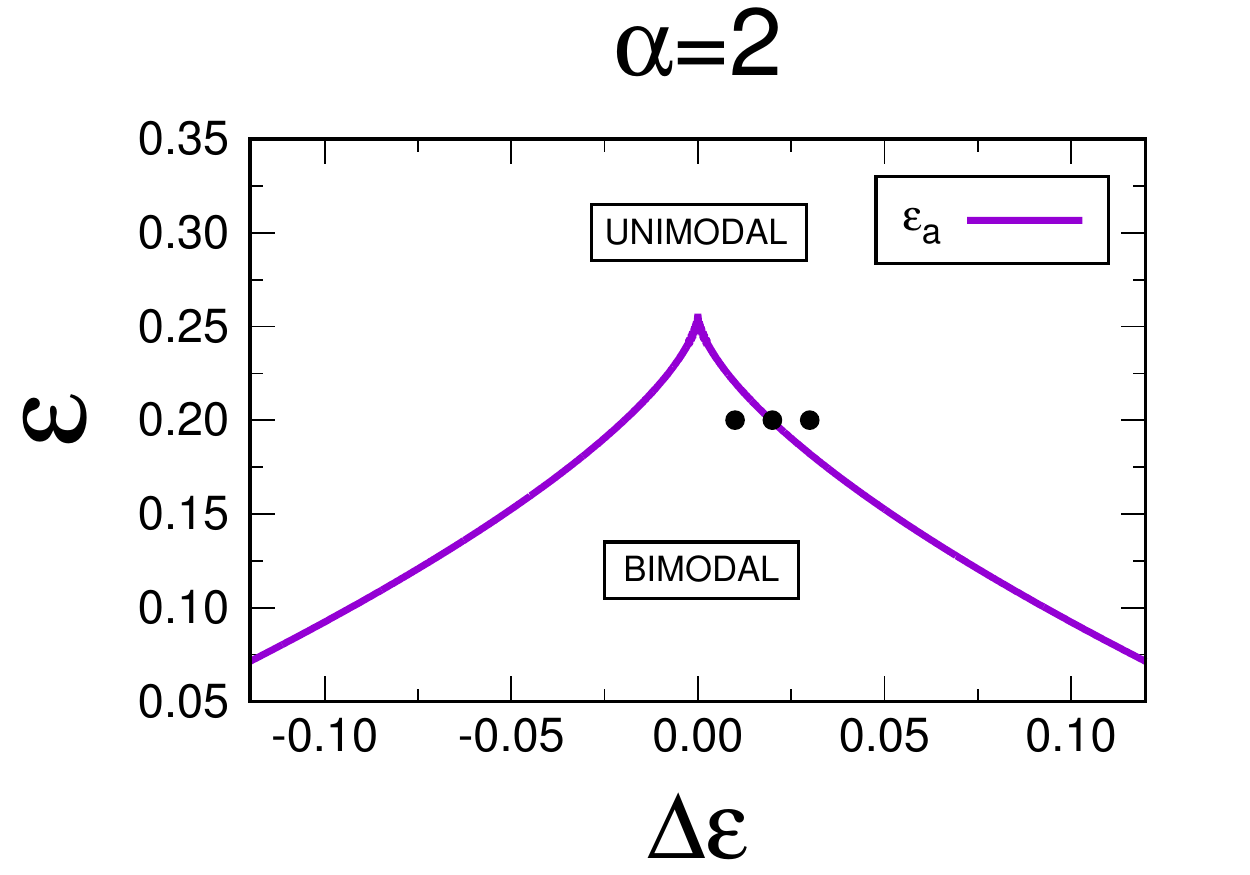}
\includegraphics[width=0.45\textwidth]{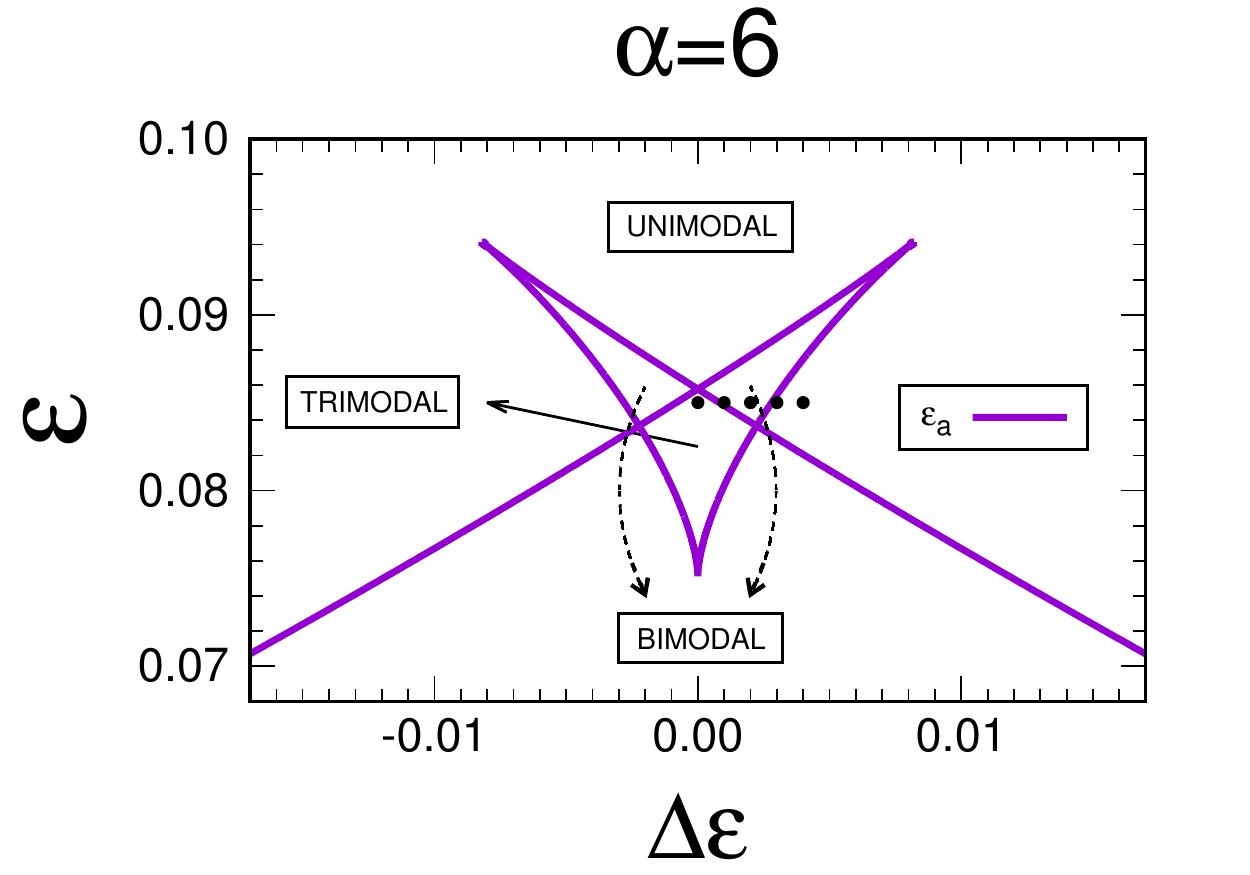}
\caption{Phase diagram of the asymmetric case $\Delta\varepsilon \neq 0$. The left panel corresponds to $\alpha=2$ and the right one to $\alpha=6$. The transition lines correspond to the expressions Eqs.(\ref{catastrophe1}). The dots are the points whose probability distributions are displayed in the bottom panels of Figs.\ref{fig:modes3} and \ref{fig:modes4}.}
\label{fig:modes2}
\end{figure*}

\subsubsection{Asymmetric case} When $a_0 \neq a_1$, there are two main effects: (i) the locations of the modes shift, and (ii) one of the members of the previous symmetrical solution pair $\pm m_{*}^{+}$ is promoted to be more likely than its corresponding partner. If the asymmetry parameter $|\Delta\varepsilon|$ is large enough, bimodality and trimodality may even be destroyed, inducing new transitions in the $(a_0,a_1,\alpha)$ or, alternatively, in the $(\varepsilon,\Delta\varepsilon,\alpha)$ space of parameters. These transitions are known as catastrophes~\citep{Cata1,Cata2}. For fixed $\alpha$, the projection $(\varepsilon_{a}, \Delta\varepsilon_{a})$ of the transition region in the plane $(\varepsilon, \Delta\varepsilon)$ must fulfill the transition conditions $F(m_{*})=0$ and $F'(m_{*})=0$, which in the approximation Eq.(\ref{drift_expand}) and in parametric representation, read
\begin{align}
\label{catastrophe1}
&\frac{\varepsilon_a }{\varepsilon_c} \approx 1 + \frac{\alpha (\alpha-5)}{2} m_{*}^2 + \frac{\alpha (\alpha-2)(\alpha-3)(\alpha-9)}{24} m_{*}^4,\\
&\frac{\Delta\varepsilon_a }{\varepsilon_c } \approx \frac{2 \alpha (\alpha-5)}{3} m_{*}^3 + \frac{\alpha (\alpha-2)(\alpha-3)(\alpha-9)}{15} m_{*}^5.\notag
\end{align}
The transition line in the plane $(\varepsilon_{a}, \Delta\varepsilon_{a})$ is obtained by elimination of the parameter $m_{*}$. As indicated in Fig.\ref{fig:modes2}, there could be several transition lines.

For $1<\alpha<5$, and when the condition Eq.(\ref{classical_second}) is fulfilled, we can disregard the terms of order $O(m^4)$ or higher in Eqs.(\ref{catastrophe1}). The resulting transition line corresponds to the well known \emph{Cusp Catastrophe}~\citep{Cata2} with geometry $\varepsilon_{c}-\varepsilon_{a} \sim \vert \Delta\varepsilon_{a} \vert^{2/3}$ (see left panel of Fig.\ref{fig:modes2}). The degenerate points $\alpha=1,5$ also display cusp-like transition curves with different geometries: $\varepsilon_{c}-\varepsilon_{a} \sim \vert \Delta\varepsilon_{a} \vert^{4/5}$ for $\alpha=5$ and $\varepsilon_{c}-\varepsilon_{a} \sim \vert \Delta\varepsilon_{a} \vert$ for $\alpha=1$. In the bottom panels of Fig.\ref{fig:modes3} we plot the probability distribution for $\alpha=2$, $\varepsilon=0.2$ as the parameter $\Delta \varepsilon$ crosses the transition line at $\Delta\varepsilon_a=0.017$ depicted in Fig.\ref{fig:modes2}.

For $\alpha>5$ one has to keep all the terms of Eqs.(\ref{catastrophe1}). In this case, trimodality, together with bimodality and unimodality, is also possible and the transition curves corresponds to the so called \emph{Butterfly Catastrophe}~\citep{Cata2}. Its geometry is similar to three cusps, whose intersections delimit the trimodal regime (see right panel of Fig.\ref{fig:modes2}, corresponding to $\alpha=6$). The two tips at the top divide unimodal from bimodal zones, while the tip at the bottom separates trimodal from bimodal. In the bottom panel of Fig.\ref{fig:modes3} we plot the probability distribution for $\alpha=6$, $\varepsilon=0.085$ as the parameter $\Delta\varepsilon$ crosses the transition lines at $\Delta\varepsilon_{a,1}=0.00082...$ and $\Delta\varepsilon_{a,2}=0.00280...$, as depicted in Fig.\ref{fig:modes2}.

\subsection{Fluctuations and N-dependence}
Fluctuations, including those due to the finite-size of the system, play a very important role in the original version of the noisy voter model, $\alpha=1$. This is mainly because, as we mentioned earlier, the transition is noise-induced and finite-size. In this case, with $\alpha=1$, and if we approach the critical point $\varepsilon_{c}(N)=1/N$ by keeping $\varepsilon \sim 1/N$ and $\Delta\varepsilon \sim 1/N$, the expression Eq.(\ref{Potential}) for the potential can be simplified to:
\begin{equation}
\label{potential_alpha1}
V(m)= \left(\frac{1}{N}-\varepsilon \right) \log (1-m^2) + \frac{\Delta\varepsilon}{2} \log \left( \frac{1-m}{1+m} \right).
\end{equation}
If we conveniently redefine the rates Eqs.(\ref{glob_rates+},\ref{glob_rates-}) such that we are always in the regime $\varepsilon \sim 1/N$ and $\Delta\varepsilon \sim 1/N$ (for example considering $h/N \equiv h^{*}$ with $h^{*}=O(1)$ being a new parameter) then the distribution $P_{\text{st}}(m)$ is $N$-independent, which is one of the interesting properties of the model with rescaled parameters widely discussed in the literature~\citep{Alfarano1,Alfarano2,Alfarano3}. We will however show that this is not the case for the non-linear version of the model and that the $N$-independence property is not a robust result of the noisy voter model.

\begin{figure*}[t]
\centering
\includegraphics[width=0.45\textwidth]{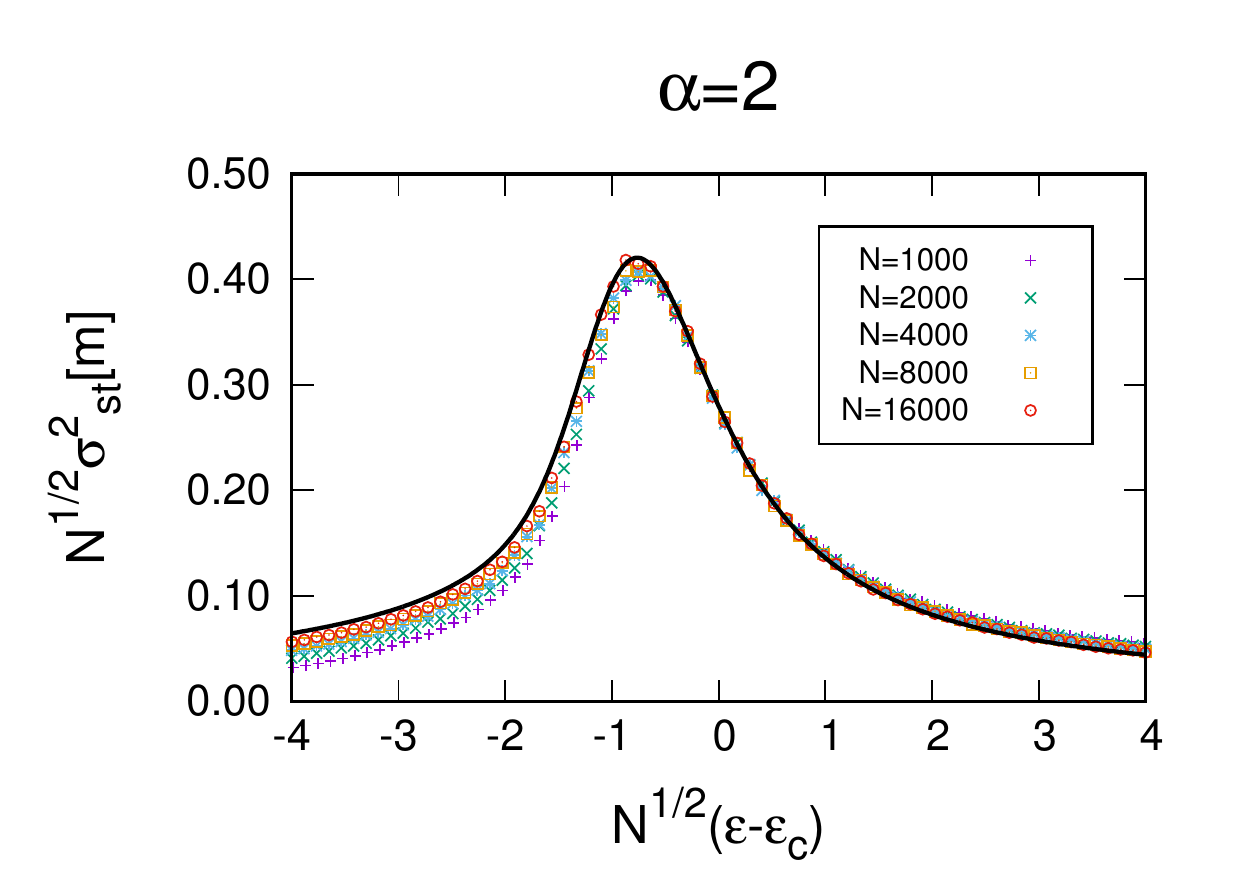}
\includegraphics[width=0.45\textwidth]{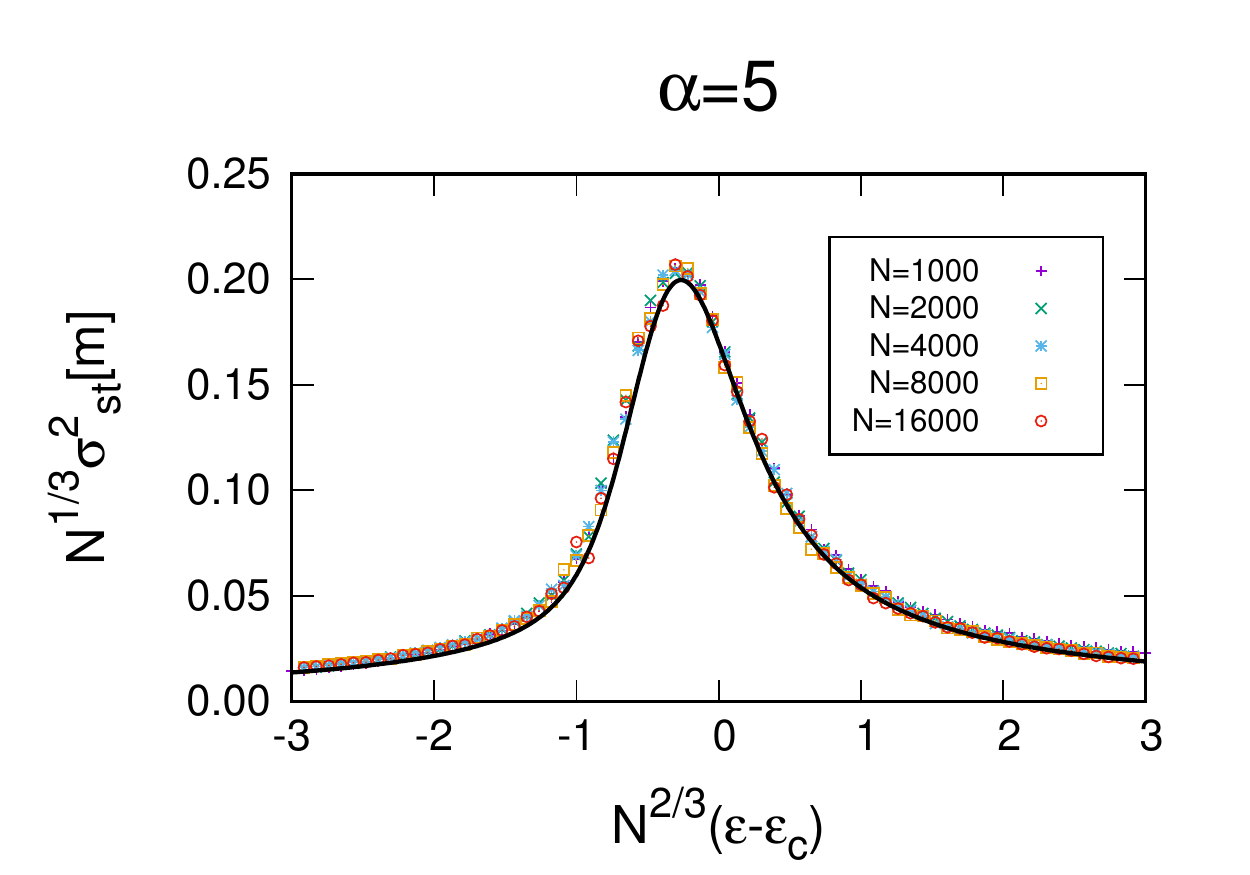}
\caption{Variance of the magnetization $\sigma_{\text{st}}^2[m] \equiv \langle m^2 \rangle_{\text{st}}-\langle \vert m \vert \rangle_{\text{st}}^2$ as a function of $\varepsilon-\varepsilon_{c}$, rescaled with the correct power of system size $N$, in order to check the scaling relations Eqs.(\ref{moment_k2},\ref{moment_k3}). Both panels are for the symmetric case $\Delta\varepsilon = 0$, the left one with $\alpha=2$ and the right one with $\alpha=5$. Dots correspond to Monte Carlo simulations of different system sizes, averaged over $10^5$ Monte Carlo steps, while the solid black line is the scaling function Eq.(\ref{phi_k},\ref{phi_k2}).}
\label{fig:fluct}
\end{figure*}

Let us expand the potential Eq.(\ref{Potential}) in power series of $m$, around the critical point $\varepsilon \sim \varepsilon_{c}$, $\Delta\varepsilon \sim 0$, to find
\begin{equation}
\label{potential_alpha}
V(m)= -c_1 m + c_2 m^2 + c_4 m^4 + c_6 m^6 +...,
\end{equation}
with coefficients
\begin{eqnarray}
\label{coeff1}
c_1 &=& \frac{2^{\alpha}}{2 \alpha} \Delta\varepsilon \equiv c \cdot \Delta\varepsilon, \\ \label{coeff2}
c_2 &=& \frac{2^{\alpha}}{2 \alpha} (\varepsilon -\varepsilon _{c}) \equiv c \cdot (\varepsilon -\varepsilon _{c}), \\ \label{coeff4}
c_4 &=& \frac{(\alpha-1)(5-\alpha)}{24}, \\ \label{coeff6}
c_6 &=& \frac{(\alpha-1)(\alpha-3)(\alpha-4)(17-\alpha)}{720}.
\end{eqnarray}
Note that we have only kept the first order of each coefficient, and only some of the coefficients. In doing so, we have neglected terms of order $1/N$, $m^2(\varepsilon -\varepsilon _{c})^2$, $m^2 \Delta\varepsilon^2$, $m^3 \Delta\varepsilon$, etc. If we now assume that $1<\alpha<5$, then we can also neglect the order $m^6$ of the potential, and an arbitrary $k-$moment of the distribution $P_{\text{st}}(m)$ will read
\begin{equation}
\label{moment_k}
\langle m^{k} \rangle_{\text{st}} = \frac{\int_{-1}^{1} dm\, m^k e^{N \left(c_1 m - c_2 m^2 - c_4 m^4 \right)}}{\int_{-1}^{1} dm\, e^{N \left(c_1 m - c_2 m^2 - c_4 m^4 \right)}}.
\end{equation}
If we make the change of variables $z=N^{1/4}m$, the leading term can be rewritten as:
\begin{equation}
\label{moment_k2}
\langle m^{k} \rangle_{\text{st}} = N^{-k/4} \phi_{k} \left[ N^{1/2} (\varepsilon -\varepsilon _{c}), N^{3/4} \Delta\varepsilon \right],
\end{equation}
with the scaling function
\begin{equation}
\label{phi_k}
\phi_{k}[x,y] = \frac{\int_{-\infty}^\infty dz\,z^k e^{c y z- c x z^2- c_{4} z^4}}{\int_{-\infty}^\infty dz\, e^{c y z- c x z^2- c_{4} z^4}}.
\end{equation}
Note that this function $\phi_{k}[x,y]$ is independent of $N$, and all the $N-$dependence is displayed explicitly in the expression Eq.(\ref{moment_k2}). The case $\alpha=5$ is a degenerate case where $c_{4}=0$, and we have to keep the order $m^6$ of the potential, this leads to a different scaling function:
\begin{equation}
\label{moment_k3}
\langle m^{k} \rangle_{\text{st}} = N^{-k/6} \phi_{k} \left[ N^{2/3} (\varepsilon -\varepsilon _{c}), N^{5/6} \Delta\varepsilon \right],
\end{equation}
with
\begin{equation}
\label{phi_k2}
\phi_{k}[x,y] = \frac{\int_{-\infty}^\infty dz\,z^k e^{c y z- c x z^2- c_{6} z^6}}{\int_{-\infty}^\infty dz\,e^{c y z- c x z^2- c_{6} z^6}}.
\end{equation}
If we write the special case $\alpha=1$ in this form, we obtain $\langle m^{k} \rangle = \phi_{k} \left[ N \varepsilon, N \Delta\varepsilon \right]$, where now the scaling function uses the expression Eq.(\ref{potential_alpha1}) of the potential with all the orders of $m$ included. In this case of $\alpha=1$, a rescaling of the parameters of the model $\varepsilon$, $\Delta\varepsilon$ in an appropriate way will make all the moments $\langle m^{k} \rangle$ $N-$independent. However, for $\alpha \neq1 $, even if we rescale the parameters of the model to be close enough to the critical point, the moments $\langle m^{k} \rangle$ vanish as $N^{-k/4}$ for $1<\alpha<5$ and as $N^{-k/6}$ for $\alpha=5$. It is also important to mention that the width of the critical region where these scaling properties hold is wider for the non-linear case $\varepsilon-\varepsilon_c \sim N^{-1/2}$ (for $1<\alpha<5$) compared to $\varepsilon-\varepsilon_c \sim N^{-1}$ for $\alpha=1$. In Fig.\ref{fig:fluct} we check the scaling form of the stationary variance of the magnetization $\sigma_\text{st}^2[m]$ in the cases $\alpha=2$ and $\alpha=5$.

\section{Complex Networks}\label{complex}
So far, we have considered the case of an all-to-all interaction, in which each node is a neighbor of every other node. Given the application of the model to problems of consensus formation in populations, it seems important to consider in detail the role the network of interactions might have on the transitions identified in the previous sections. In this way we are able to go beyond the global coupling approach and consider in some detail the effect of the locality of the interactions.

There are several theoretical approaches under which the network structure can be considered~\citep{Gleeson1,Gleeson2,Carro2,Peralta,Redner,Pugliese,Alfarano3}. In this work we focus on a particular version of the so-called {\slshape pair approximation} as developed in~\citep{Vazquez1,Vazquez2}. In this approach, one defines the network related variables ${\bf n}_k$ and $\rho$ as, respectively, the number of nodes with degree $k$ in state $1$ (not to be confused with $n_i$, which is the node state variable) and the density of active links, i.e. the ratio between the number of links connecting
nodes in different states and the total number of links $\mu N/2$. Additionally, it is also convenient to define the variable $n_L = \sum_{i=1}^{N} k_{i} n_{i}$, which is the number of links coming out of nodes in state $1$. The intensive versions of the variables ${\bf n}_k$ and $n_L$ are $m_k=2 {\bf n}_k/N_k-1$ and $m_L = 2 n_L/N\mu-1$ (also known as {\slshape link magnetization}). We follow closely the method of Vazquez and Eguiluz~\citep{Vazquez2} in their study of the linear noiseless voter model adapted to the presence of the noise term and non-linear rates of our model, and we refer to that work for a more detailed explanation of the approach. Similar, but not identical, treatments have been developed by Diakonova et al.~\citep{Diakonova} in their study of the linear noisy voter model, and by J\polhk{e}drzejewski\cite{Jedrzejewski} and by Min and San Miguel~\citep{Min} in particular versions of nonlinear voter models.

For simplicity, we will restrict ourselves to the symmetric case $\Delta\varepsilon = 0$ and $a_{0} = a_{1} \equiv a$, such that $m_k=m_L=0$ is always a fixed point of the dynamics. We start by finding an approximate mean-field time evolution equation for $\rho$:
\begin{align}
\label{rho_evolution1}
\frac{d \rho}{d t} &= \sum_{k} \frac{P_{k}}{2} \sum_{q=0}^{k} P_{0}(k,q) \left( a + h \left(\frac{q}{k} \right)^\alpha \right) \frac{\Delta \rho_{k,q}}{\Delta t}\\
&-\sum_{k} \frac{P_{k}}{2} \sum_{q=0}^{k} P_{1}(k,q) \left( a + h \left( \frac{k-q}{k} \right)^\alpha \right) \frac{\Delta \rho_{k,q}}{\Delta t}.\notag
\end{align}
Here $P_{0/1}(k,q)$ is the probability of selecting a node within the population in state $0/1$ with degree $k$ and that has $q$ neighbors in state $1$, $\Delta \rho_{k,q}=2(k-2q)/\mu N$ is the change in $\rho$ when a node with $(k,q)$ changes from $0$ to $1$ (being $-\Delta \rho_{k,q}$ the corresponding change in $\rho$ when a $(k,q)$ node changes from $1$ to $0$), and $\Delta t = 1/N$ is the elementary time increment in Monte Carlo steps\citep{f1}. In this way, the right-hand side of Eq.(\ref{rho_evolution1}) is simply the probability of selecting nodes $(0/1,k,q)$, times the rate at which those nodes change state, times the corresponding change in $\rho$, summed over all the possible values of $k,q$. The so called pair approximation assumes $P_{0/1}(k,q)$ to be binomial, with a single event probability $p_0=\rho$ and $p_1=1-\rho$, i.e. $P_{0}(k,q) \approx {{k}\choose{q}} \rho^q (1-\rho)^{k-q}$, $P_{1}(k,q) \approx {{k}\choose{q}} (1-\rho)^q \rho^{k-q}$. Under this approximation, Eq.(\ref{rho_evolution1}) reduces to
\begin{align}
\label{rho_evolution2}
\frac{d \rho}{d t} = 2 a (1-2\rho)+ \frac{2 h}{\mu} \sum_{k} P_{k} \left\langle (k-2q) \left(\frac{q}{k}\right)^{\alpha} \right\rangle_{0},
\end{align}
where $\langle... \rangle_{0/1}$ is an average over $P_{0/1}(k,q)$. In the case of directed networks\citep{Konstantin} the density of $0\rightarrow1$ links $\rho_{01}$ is not the same as the density of  $1\rightarrow0$ links $\rho_{10}$, and one would need to obtain an evolution equation for each. While such a generalization is beyond the scope of this contribution, we point the interested reader to Ref.~\citep{directed}, where the necessary steps are described.

\begin{figure}[h]
\centering
\includegraphics[width=0.45\textwidth]{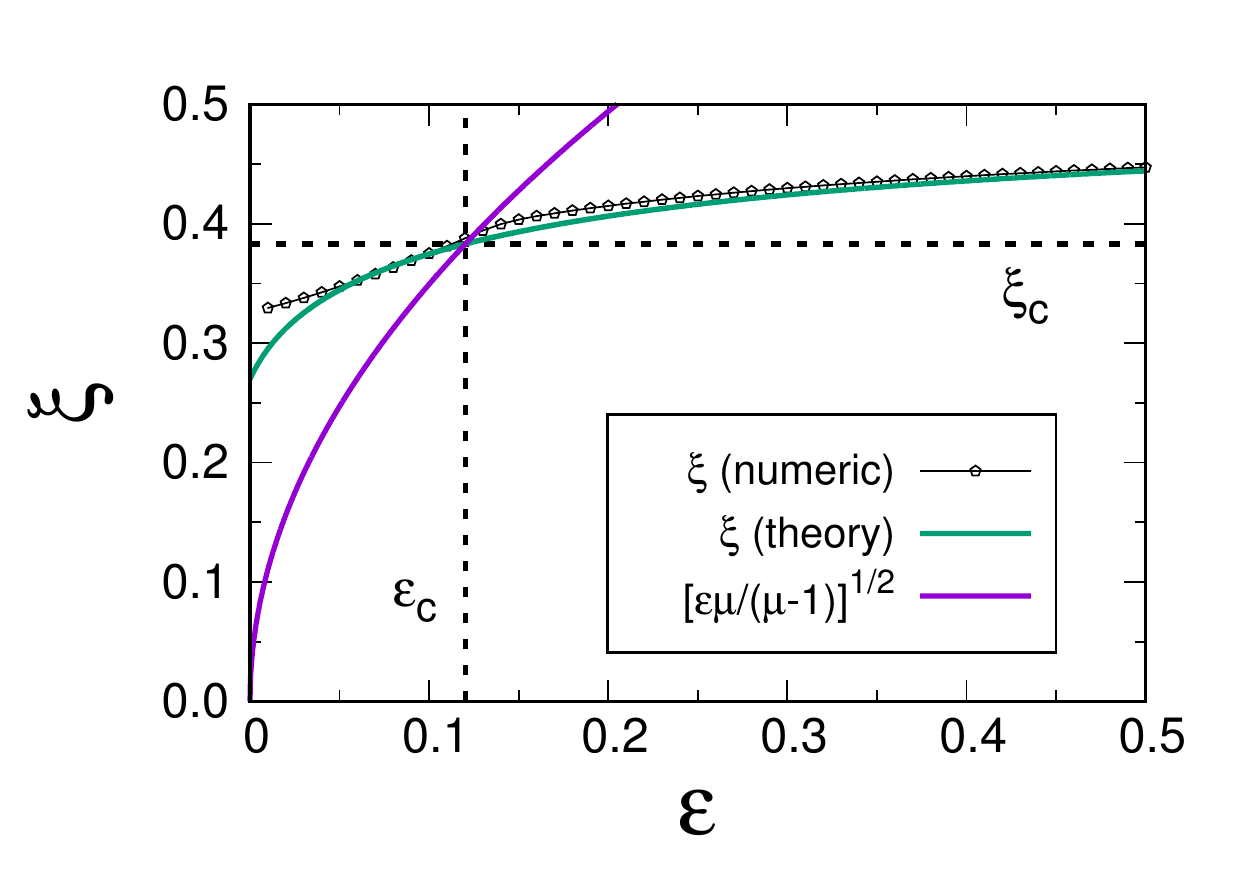}
\caption{Stationary solution $\rho=\xi$ from Eq.(\ref{xi}), corresponding to $\alpha=2$, as a function of $\varepsilon$ for $\mu=5.54$, $\mu_{-1}=0.35$ and $\omega_1=0.30$, $\omega_2=0.64$, $\omega_3=3.24$ (green line). The black line-point are the numerical results of $\xi(\text{numeric}) \equiv \langle \rho \rangle_{\text{st}}/\langle 1 - m^2\rangle_{\text{st}}$ for an ensemble of $100$ scale free networks with the specified degree moments. In purple $\sqrt{\dfrac{\varepsilon \mu}{\mu-1}}$, the crossing of the two curves is the critical point $\varepsilon_{c}=0.120$, $\xi_{c}=0.383$.}
\label{fig:xi}
\end{figure}

Note that Eq.(\ref{rho_evolution2}) is applicable for an arbitrary $\alpha$ and that, in particular, for integer $\alpha$, the right-hand side contains powers of $\rho$ up to order $\alpha+1$. It also depends on the mean degree $\mu$ and the first negative moments of the degree distribution $\mu_{-1}$,..., $\mu_{-\alpha+1}$, e.g. for $\alpha=1$ it is~\citep{Diakonova}
\begin{align}
\label{rho_evolution3a}
\frac{d \rho}{d t} =-4\rho^2\frac{\mu-1}{\mu}+2\rho\left(\frac{\mu-2}{\mu}-2a\right)+2a,
\end{align}
and for $\alpha=2$,
\begin{align}
\label{rho_evolution3}
\frac{d \rho}{d t} = 2 a (1-2\rho)+ \frac{2 h \rho}{\mu} \left( \omega_1 + \omega_2 \rho - 2 \omega_3 \rho^2 \right),
\end{align}
with $\omega_1=1-2\mu_{-1}$, $\omega_2=\mu-7+6\mu_{-1}$ and $\omega_3=\mu-3+2\mu_{-1}$ (as no node of the network is allowed to be isolated, the degree distribution satisfies $P_0=0$ and the $m$-moment $\mu_m$ is defined also for negative $m$, as required in the previous formulas). Although we have not been able to provide a rigorous proof, for all analyzed cases it turns out that the stationary state $\dfrac{d \rho}{d t}=0$ admits a physical (stable) solution $\rho=\xi\in[0,1]$. For $\alpha=2$, $\xi$ is the root of the cubic equation
\begin{align}
\label{xi}
\varepsilon (1-2\xi)+ \frac{\xi}{\mu} \left( \omega_1 + \omega_2 \xi - 2 \omega_3 \xi^2 \right)=0,
\end{align}
satisfying $\xi\in[0,1]$. See Fig.\ref{fig:xi} for an example of the dependence of $\xi$ with $\varepsilon=a/h$. It is possible to recover the all-to-all limit by taking a degree distribution $P_k=\delta(k-\mu)$ such that all averages can be written as $\mu_m=\mu^m$, and taking the limit $\mu\to\infty$. In this limiting case, it is found that $\xi=1/2$.

\begin{figure}[h]
\centering
\includegraphics[width=0.45\textwidth]{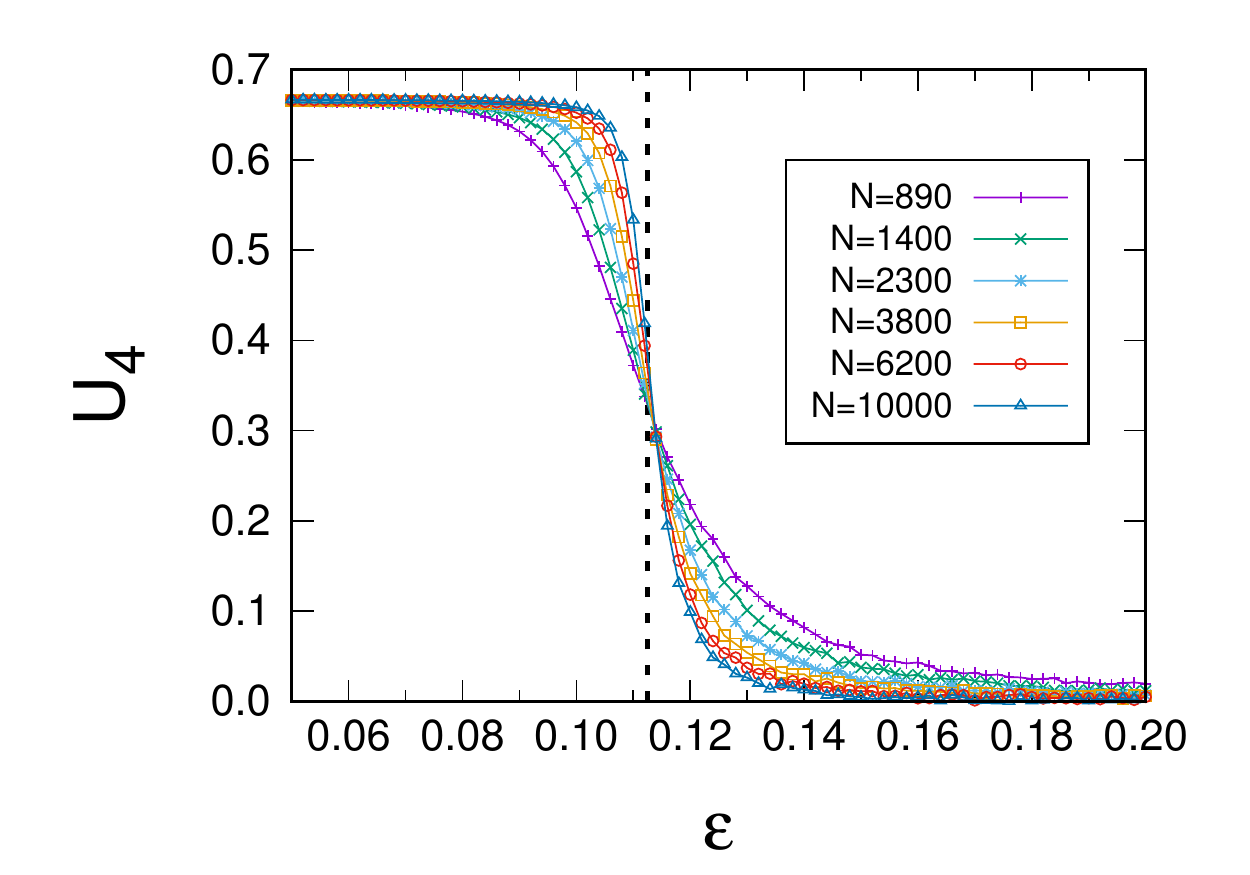}
\caption{Binder cumulant as a function of $\varepsilon$ for a 5-regular network. Dots (joined by lines as a guide to the eye) correspond to Monte Carlo simulations of different system sizes, averaged over $10^6$ Monte Carlo steps, while the vertical dashed black line is the prediction $\varepsilon_c=0.1125$ of the pair-approximation.}
\label{fig:fluct_net1_reg}
\end{figure}

\begin{figure}[h]
\centering
\includegraphics[width=0.45\textwidth]{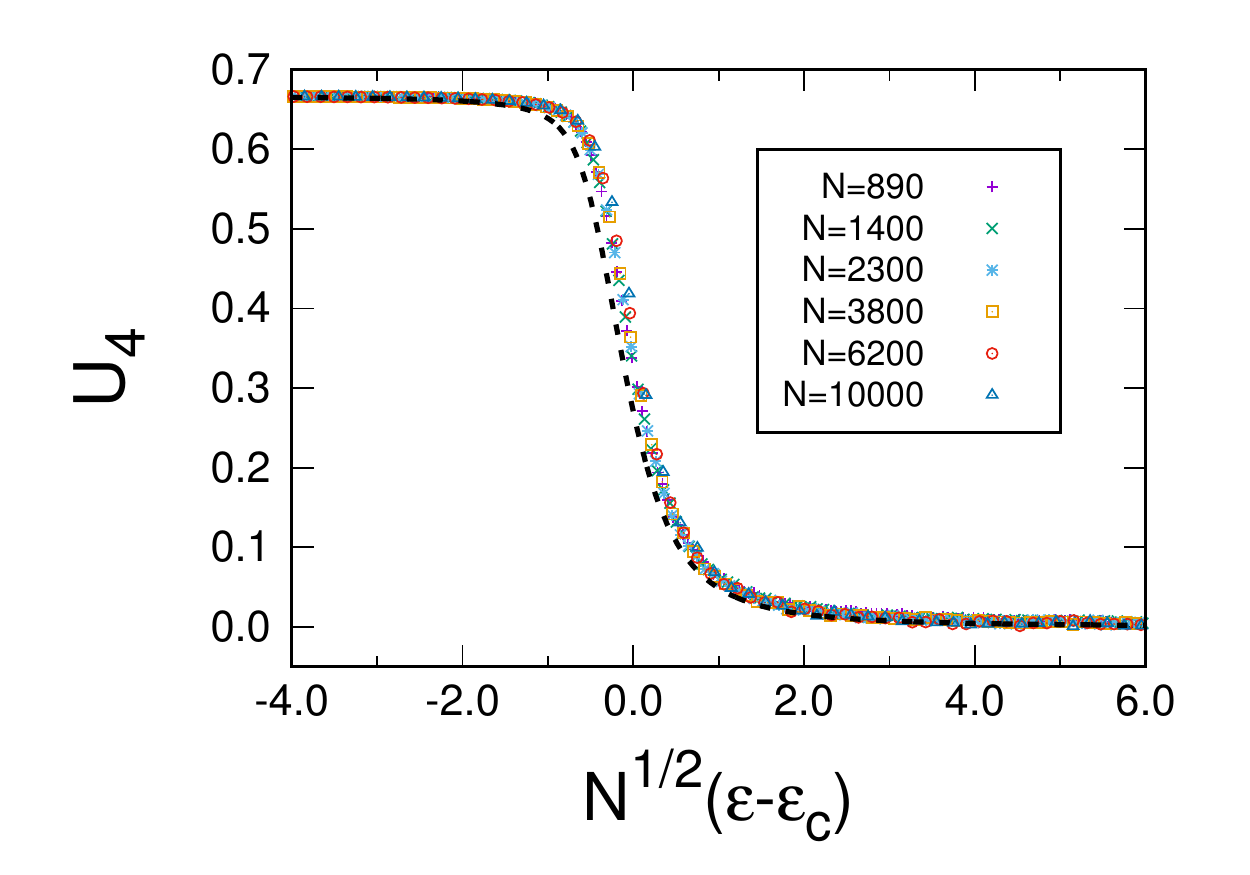}
\caption{The same data of Fig.\ref{fig:fluct_net1_reg} are plotted versus the rescaled variable $N^{1/2}(\varepsilon-\varepsilon_c)$. The good collapse of the data validates the proposed scaling law $U_4(\varepsilon,N)=u(N^{1/2}(\varepsilon-\varepsilon_c))$, being $u(x)$ the scaling function. The theoretical prediction (dotted line) is obtained from Eq.(\ref{phi_k}) using the coefficients $c=1.602$ and $c_{4}=0.093$.}
\label{fig:fluct_net2_reg}
\end{figure}

\begin{figure}[h]
\centering
\includegraphics[width=0.45\textwidth]{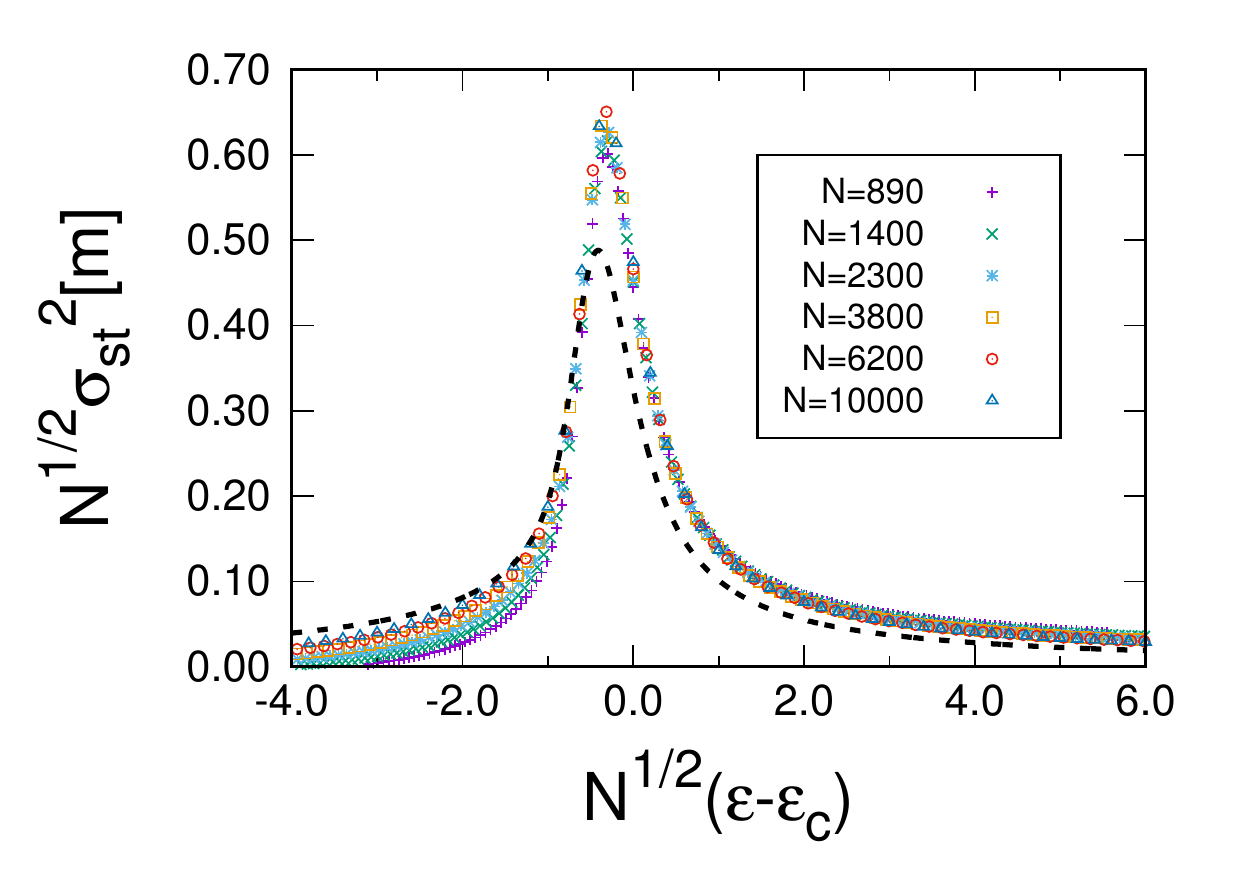}
\caption{Rescaled variance of the magnetization as a function of $N^{1/2}(\varepsilon-\varepsilon_c)$ for a 5-regular network. Dots correspond to Monte Carlo simulations of different system sizes, averaged over $10^6$ Monte Carlo steps, while the dashed black line is the prediction of the pair-approximation.}
\label{fig:fluct_net3_reg}
\end{figure}

Next, we derive a master equation for the link-magnetization variable $m_L$. A single update $0\to 1$ of a node with degree $k$ implies a change $m_L \rightarrow m_L +\Delta_{k}$, with $\Delta_{k} = \frac{2k}{\mu N}$, whereas an update $1\to 0$ of a node with degree $k$ changes $m_L \rightarrow m_L -\Delta_{k}$. The master equation for $P(m_L,t)$ then reads
\begin{equation}
\label{2Master_equation}
\frac{\partial P(m_L,t)}{\partial t} = \sum_{k} \left(E_{m_L}^{\Delta_k}-1 \right) \left[ \pi_{k}^{-} P \right] + \left(E_{m_L}^{-\Delta_k}-1 \right) \left[ \pi_{k}^{+} P \right],
\end{equation}
where $E_{m_L}^\ell$ is the step operator acting on an arbitrary function $g(m_L)$ as $E_{m_L}^\ell[g(m_L)]=g(m_L+\ell)$, and $\pi_{k}^{\pm}$ are the rates of the proposed processes in the $m_L$ variable:
\begin{eqnarray}
\label{2glob_rates+}
\pi_{k}^{+} &=& (N_k-{\bf n}_k) \left( a + h \left\langle \left( \frac{q}{k} \right)^{\alpha} \right\rangle_{0} \right),\\
\label{2glob_rates-}
\pi_{k}^{-} &=& {\bf n}_k \left( a + h \left\langle \left(\frac{k-q}{k} \right)^{\alpha} \right\rangle_{1} \right).
\end{eqnarray}
The single event probabilities $p_{0/1}$ are different as $m_L \neq 0$ during the dynamical evolution. They are calculated as the ratio between the number of active links and the number of links coming out from nodes in state $0/1$, i.e. $p_{0}=\frac{\rho}{1-m_L}$ and $1-p_{1}=\frac{\rho}{1+m_L}$. If we expand the master equation in powers of $\Delta_k$ to second order we find
\begin{equation}
\label{2Fokker-Planck_equation}
\frac{\partial P(m_L,t)}{\partial t} = - \frac{\partial}{\partial m_L} \left[ F_L(m_L) P \right]+ \frac{1}{N} \frac{\partial^2}{\partial m_{L}^2} \left[ D_L(m_L) P \right],
\end{equation}
with drift $F_L = \sum_{k} \dfrac{2k}{\mu N} \left[ \pi_{k}^{+} - \pi_{k}^{-} \right]$ and diffusion $D_L = \sum_{k} \dfrac{2k^2}{\mu^2 N} \left[ \pi_{k}^{+} + \pi_{k}^{-} \right]$.

In order for Eq.(\ref{2Master_equation}) to be consistent, $\pi_{k}^{\pm}$ should depend only on $m_L$. Following~\citep{Redner, Peralta,Vazquez2} we will make an approximation based on an adiabatic elimination that allows us to relate variables $m_k\approx m_L$, $\rho \approx \xi(1-m_L^2)$ as functions of $m_L$ only. Performing the averages on Eqs.(\ref{2glob_rates+},\ref{2glob_rates-}) using the known moments of the binomial distributions $P_{0/1}(k,q)$ and replacing the above mentioned relations resulting from the adiabatic approximation we obtain specific formulas for the drift and diffusion in terms of the $m_L$ variable. For example, if we take $\alpha=2$ we obtain
\begin{align}
\label{2drift}
&F_L = - 2 a m_L + 2 h \xi^2 \frac{\mu-1}{\mu} m_L(1-m_L^2),\\
\label{2diffusion}
&D_L = \frac{\mu_2}{\mu^2} \left(2 a + 2 h \xi \frac{\mu+(\mu_2-\mu)\xi}{\mu_2}(1-m_L^2) \right),
\end{align}
which coincide with Eqs.(\ref{drift},\ref{diffusion}) in the all-to-all limit $\mu_{m}=\mu^{m}$, $\mu \rightarrow \infty$, $\xi=1/2$. It is now straightforward to repeat the analysis of section III with those new drift and diffusion terms Eqs.(\ref{2drift},\ref{2diffusion}) or their equivalent for other values of $\alpha$. The critical point is determined by the condition $\varepsilon_{c}=\dfrac{\mu-1}{\mu} \xi_c^2$ where the dependence of $\xi_c$ with $\varepsilon_c$ is given by Eq.(\ref{xi}) or its equivalent for other values of $\alpha$. Note that $\varepsilon_c$ is always smaller, for finite $\mu$, than the all-to-all result $\varepsilon_{c}=1/4$ (see Fig.\ref{fig:xi}).  For the particular case of a $15$-regular lattice, we find that the tricritical point moves to a larger value $\alpha=6.14$, with corresponding $\varepsilon=0.084$, and the trimodal region shrinks with respect to the one found in the all-to-all setup (see Fig.\ref{fig:modes1}). For $z$-regular lattices with $z\le 5$ the trimodal region disappears altogether. In general, multiple stability and tricritical points are also possible in the case of other complex networks and we leave for future work a detailed analysis of the phase diagram for different network types.

We now compare the theoretical results with numerical simulations of the stochastic process as defined by the individual rates Eqs.(\ref{ind_rates+},\ref{ind_rates-}). We consider a quadratic interaction $\alpha=2$ and two network configurations: (i) a $z$-regular random network, where each node is randomly connected to exactly $z$ neighbors and hence $\mu=z$ and $\mu_{m}=\mu^{m}$; and (ii) a scale-free network, with a power-law degree distribution $P_{k} \sim k^{-\lambda}$. Both networks have been generated with the configuration model as detailed in~\citep{Romualdo}. 

In the $z$-regular network, for which we take $z=5$, the previous analysis leads to $\varepsilon_{c}=0.1125$, $\xi_{c}=0.3899$ and coefficients of the potential [see Eq.(\ref{potential_alpha}) and Eq.(\ref{coeff2})] $c=1.602$, $c_4=0.093$, while the all-to-all solution is $\varepsilon_{c}=0.25$, $\xi_{c}=0.5$ and $c=1$, $c_4=0.125$. To compare the predictions of the pair approximation with the numerical simulations in the steady state, we have chosen two particular combinations of moments: the Binder cumulant~\citep{Binder,Toral-Colet:2014}, defined as $U_4=1-\dfrac{\langle m^4 \rangle}{3 \langle m^2 \rangle^2}$, and the variance $\sigma^2[m]$. As shown in Fig.\ref{fig:fluct_net1_reg}, the theory agrees very well with the simulations. According to the finite-size scaling relation Eq.(\ref{moment_k2}) for $\Delta\varepsilon=0$, the dependence on system size is $U_4(\varepsilon,N)=u(N^{1/2}(\varepsilon-\varepsilon_c))$, being $u(x)$ the scaling function. This means that the cumulant curves for different values of $N$ intersect at the critical point of the infinite system $\varepsilon_c$. From the numerical data we obtain $\varepsilon_{c}=0.114\pm0.002$ which is compatible with the theoretical prediction. A plot of $U_4(\varepsilon,N)$ vs $N^{1/2}(\varepsilon-\varepsilon_c)$ (see Fig.\ref{fig:fluct_net2_reg}) indicates a good data collapse in a single function $u(x)$. For the variance we obtain the scaling law $\sigma^2[m](\varepsilon,N)=N^{-1/2}v(N^{1/2}(\varepsilon-\varepsilon_c))$, being $v(x)$ the scaling function, which is checked in Fig.\ref{fig:fluct_net3_reg}. The comparison in Figs.\ref{fig:fluct_net2_reg},\ref{fig:fluct_net3_reg} indicates that the scaling functions $u(x)$ and $v(x)$ obtained analytically capture reasonably well the overall shape of the numerical results.

\begin{figure}[h]
\centering
\includegraphics[width=0.45\textwidth]{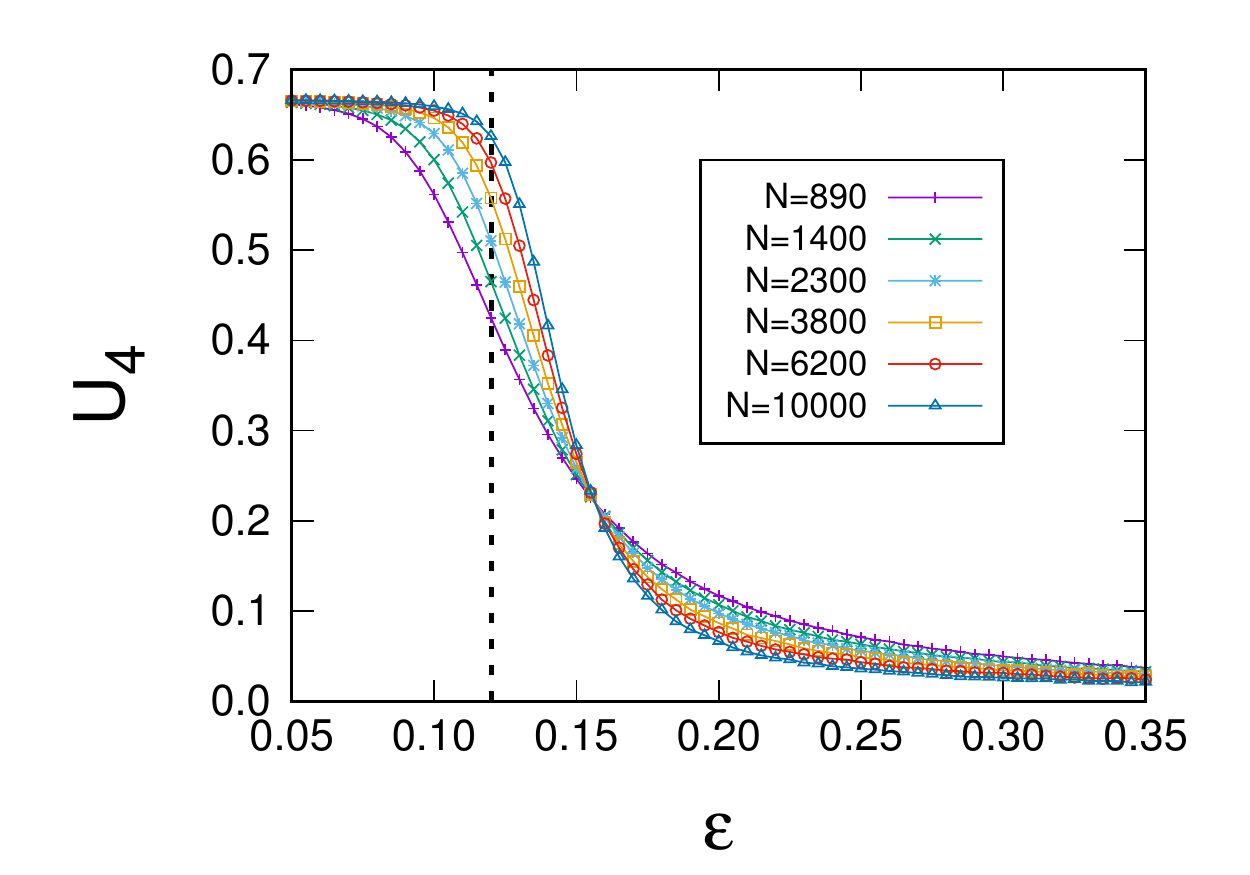}
\caption{Binder cumulant as a function of $\varepsilon$ for a scale-free network with $\lambda=2.3$. Dots (joined by lines as a guide to the eye) correspond to Monte Carlo simulations of different system sizes, averaged over $10^5$ Monte Carlo steps and an ensemble of $100$ networks, while the dashed black line is the prediction $\varepsilon_c=0.120$ of the pair approximation.}
\label{fig:fluct_net1}
\end{figure}

\begin{figure}[h]
\centering
\includegraphics[width=0.45\textwidth]{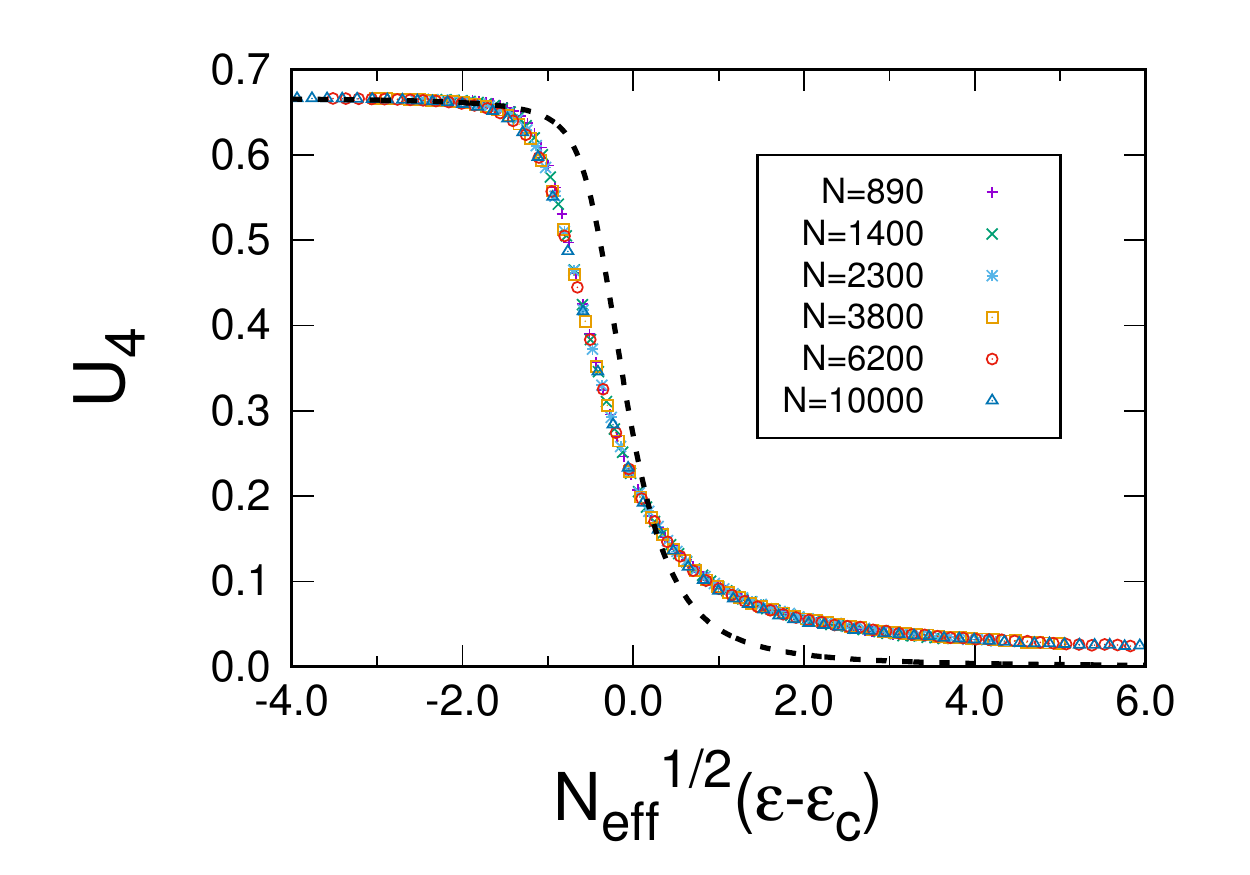}
\caption{The same data of Fig.\ref{fig:fluct_net1} are plotted versus the rescaled variable $N_\text{eff}(\varepsilon-\varepsilon_c)^{1/2}$. The good collapse of the data validates the proposed scaling law $U_4(\varepsilon,N)=u(N_\text{eff}^{1/2}(\varepsilon-\varepsilon_c))$, being $u(x)$ the scaling function. The theoretical prediction (dotted line) is obtained from Eq.(\ref{phi_k}) using instead of $c$, $c_4$ the rescaled coefficients $\bar{c}=1.84$ and $\bar{c}_{4}=0.11$.}
\label{fig:fluct_net2}
\end{figure}

\begin{figure}[h]
\centering
\includegraphics[width=0.45\textwidth]{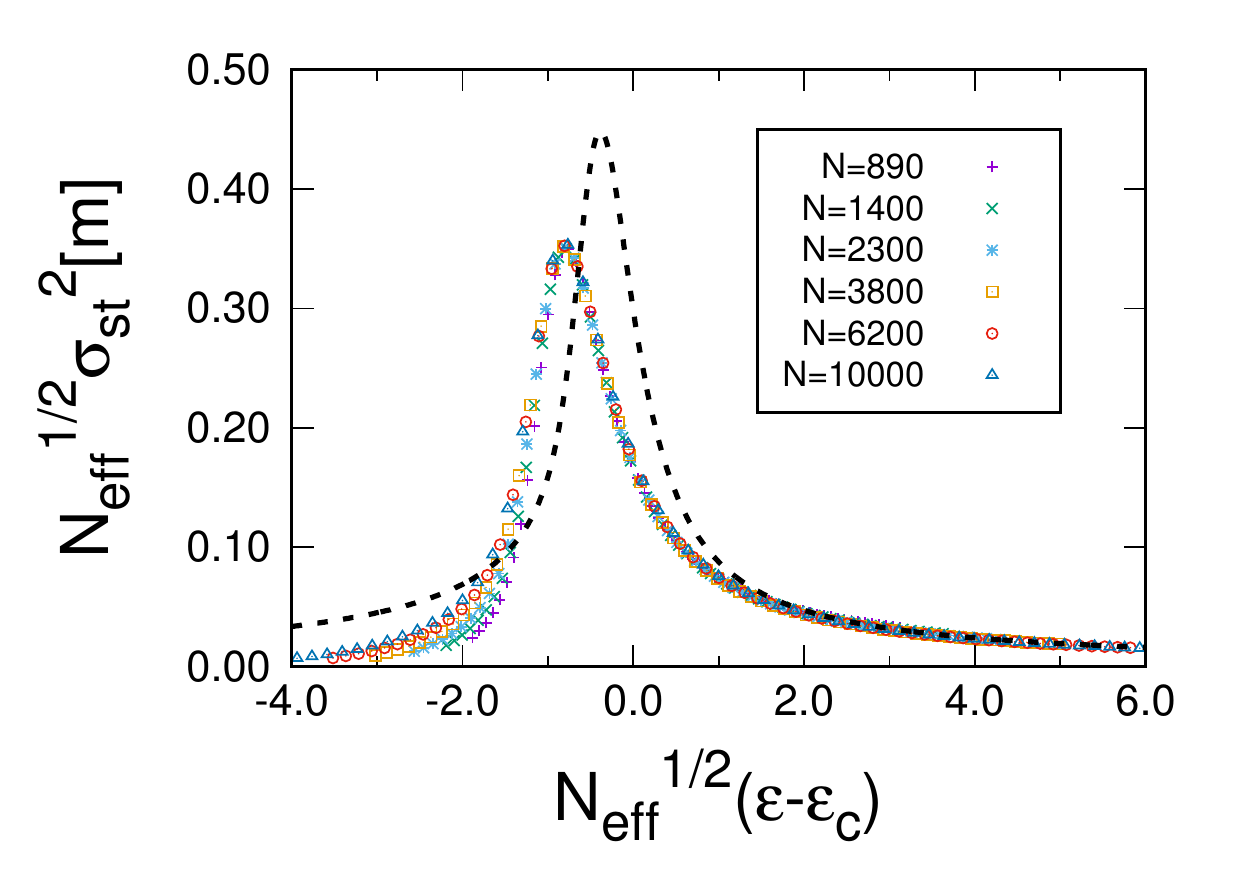}
\caption{Rescaled variance of the magnetization as a function of $\varepsilon$ in the same case than in Fig.\ref{fig:fluct_net1}, showing the validity of the scaling law $\sigma^2[m](\varepsilon,N)=N_\text{eff}^{-1/2}v(N_\text{eff}^{1/2}(\varepsilon-\varepsilon_c))$. The theoretical prediction (dotted line) is obtained from Eq.(\ref{phi_k}) using instead of $c$, $c_4$ the rescaled coefficients $\bar{c}=1.84$ and $\bar{c}_{4}=0.11$.}
\label{fig:fluct_net3}
\end{figure}

 In the scale-free network, we have chosen $\lambda=2.3$, and measured directly $\mu \approx 5.54$, $\mu_{-1} \approx 0.35$, while the second moment depends on system size $N$ as $\mu_2(N)\sim N^b$ with $b=0.64$. The theoretical results are $\varepsilon_{c}=0.120$ and $\xi_{c}=0.383$. The analysis of the Binder cumulant obtained numerically (see Fig.\ref{fig:fluct_net1}) yields $\varepsilon_c=0.157$, while the all-to-all prediction is $\varepsilon_c=0.25$. The finite-size scaling of the scale-free network is somehow tricky. In our theoretical analysis the critical point $\varepsilon_{c}$ depends only on $\mu$ and $\mu_{-1}$, which tend to finite values as $N\to\infty$, but the second moment $\mu_2$ diverges as $N^b$. Under the assumption $\mu_2\gg\mu$, which is true in general for scale-free networks, the diffusion scales as $D_L \propto \mu^2/\mu_2$ [see, for example, Eq.(\ref{2diffusion}) for $\alpha=2$], which eventually leads to coefficients $c \propto \mu^2/\mu_2$ and $c_4 \propto \mu^2/\mu_2$. One can still keep a scaling law as in Eq.(\ref{moment_k2}) if one replaces the size $N$ by an effective system size defined as $N_{\text{eff}}=N\mu^2/\mu_2$, in such a way that the dependence of $c$ and $c_4$ on $\mu_2$ is absorbed by $N_{\text{eff}}$ and the scaling law becomes $U_4(\varepsilon,N)=u(N_\text{eff}^{1/2}(\varepsilon-\varepsilon_c))$, being $u(x)$ the scaling function. This scaling law is confirmed by the data collapse presented in Fig.\ref{fig:fluct_net2}. The theoretical scaling function obtained from Eq.(\ref{moment_k2}) replacing $c$ and $c_4$ by their rescaled versions $\bar{c}=c\mu_2/\mu^2=1.84$ and $\bar{c}_{4}=c_4\mu_2/\mu^2=0.11$ yields a reasonable qualitative agreement with the numerical data. A similar scaling law holds for the variance $\sigma^2[m](\varepsilon,N)=N_\text{eff}^{-1/2}v(N_\text{eff}^{1/2}(\varepsilon-\varepsilon_c))$, the validity of which is checked in Fig.\ref{fig:fluct_net3}. While the numerical results indeed follow the proposed scaling law, the theory can only offer a qualitative agreement for the scaling function. 
 
\begin{figure*}[h]
\centering
\includegraphics[width=0.32\textwidth]{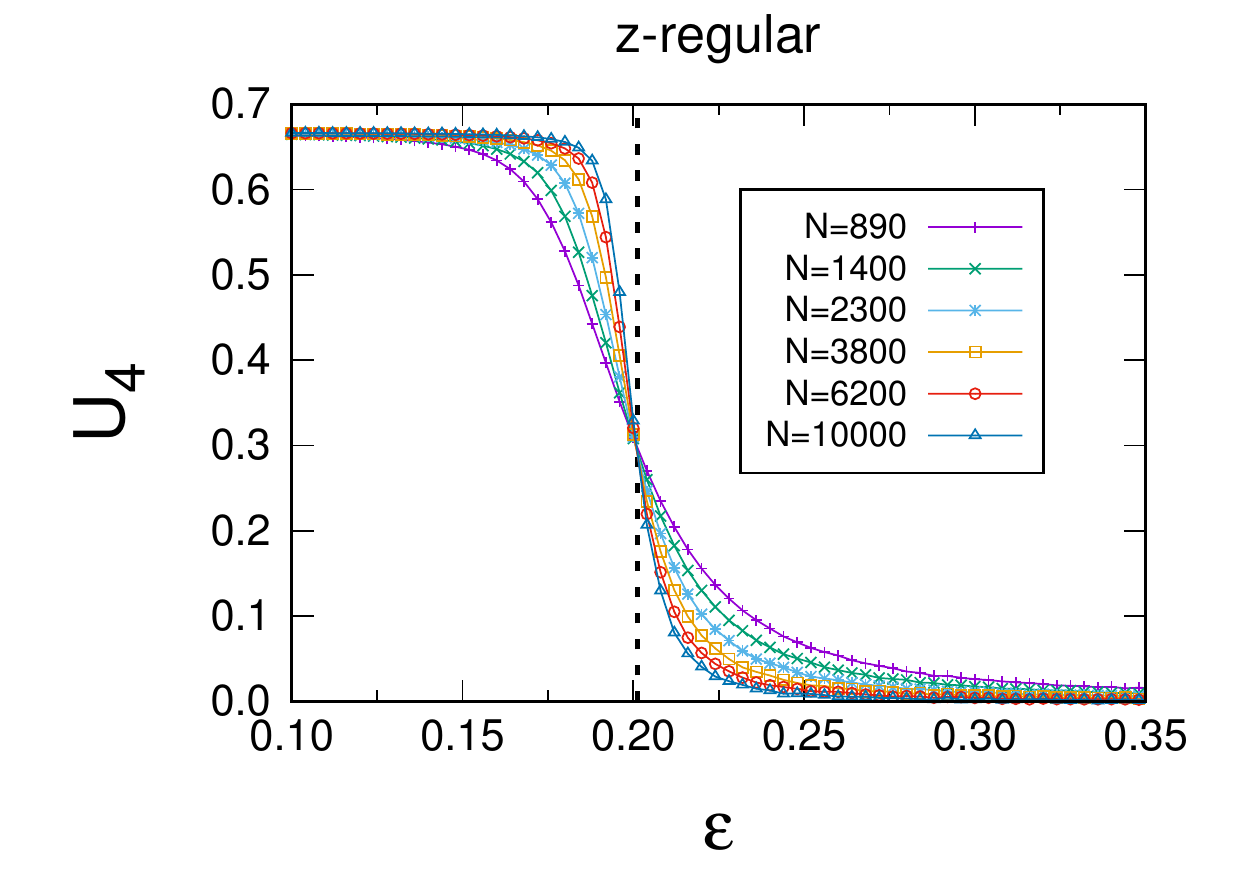}
\includegraphics[width=0.32\textwidth]{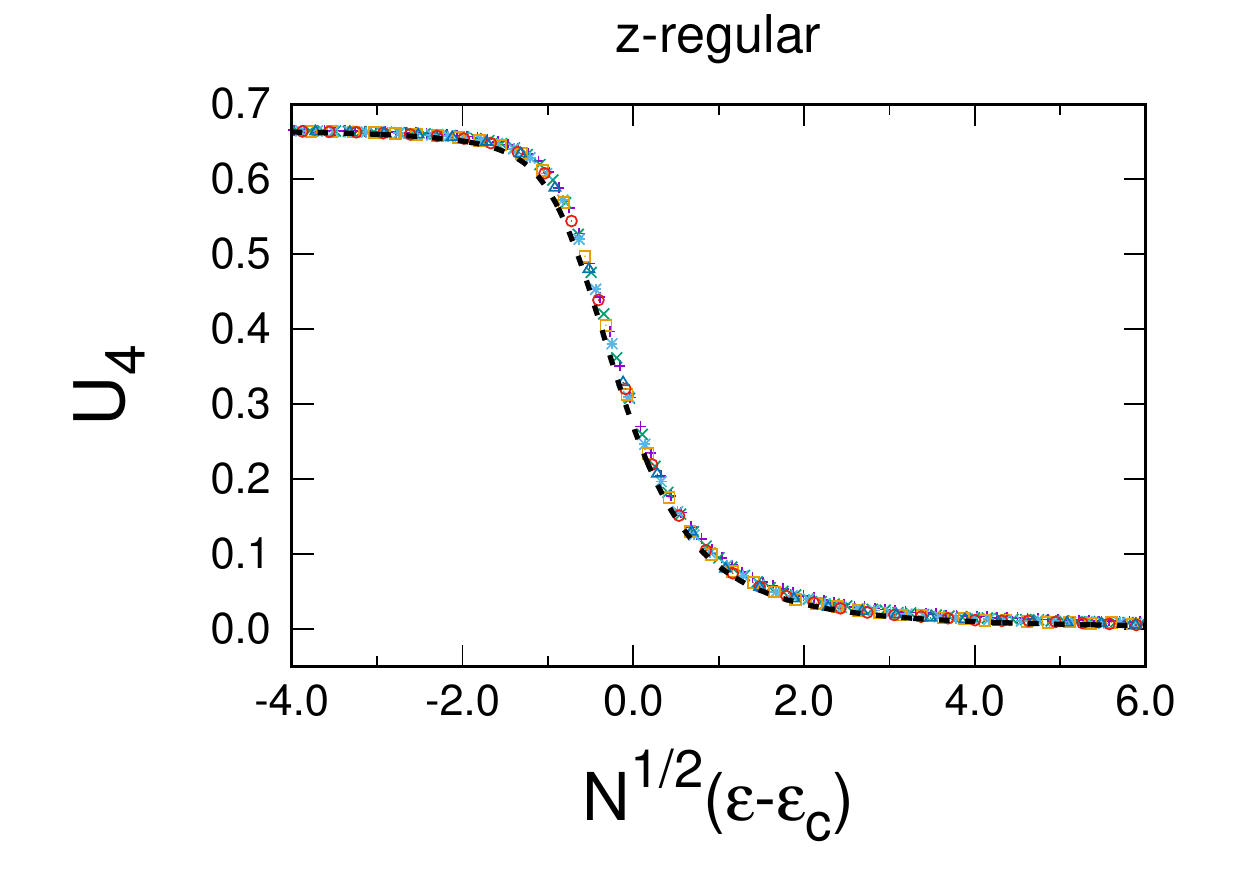}
\includegraphics[width=0.32\textwidth]{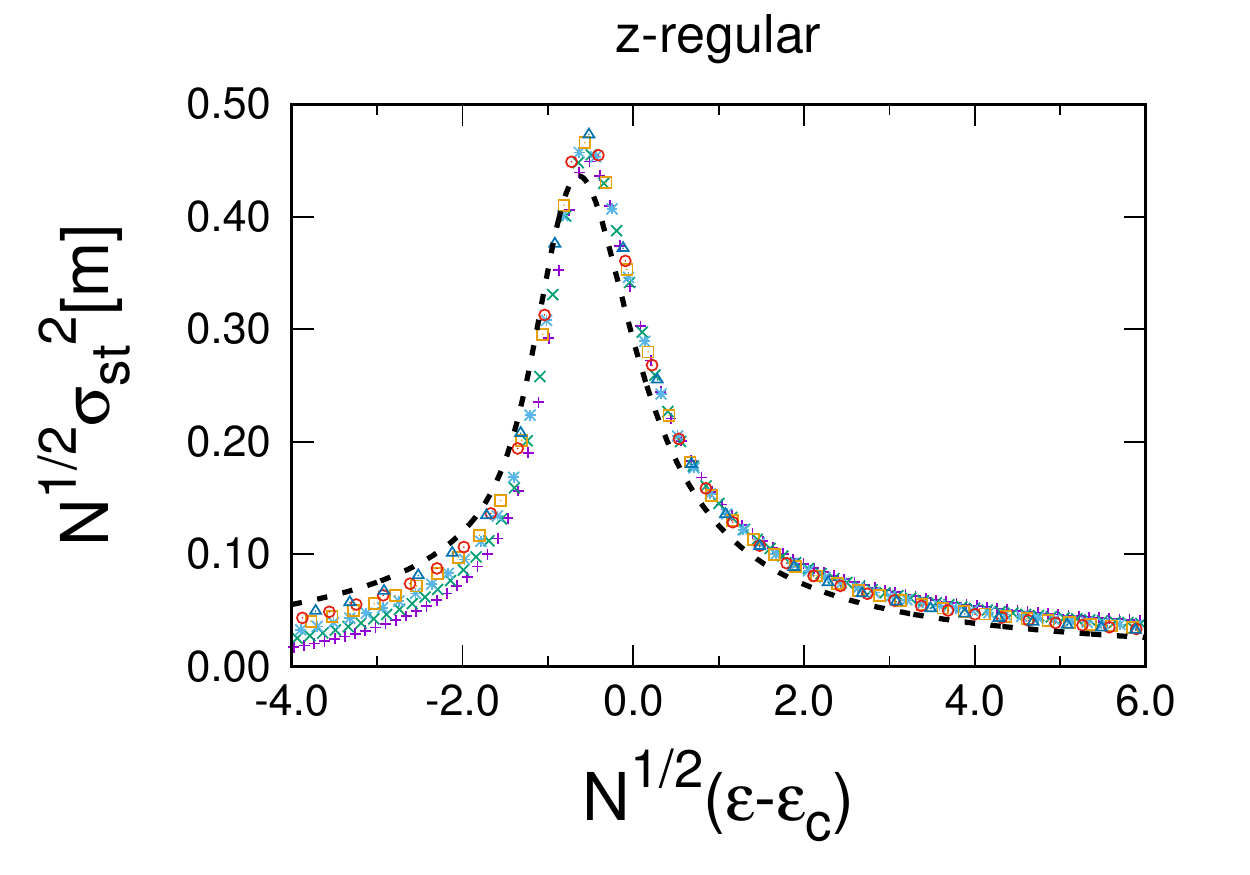}

\includegraphics[width=0.32\textwidth]{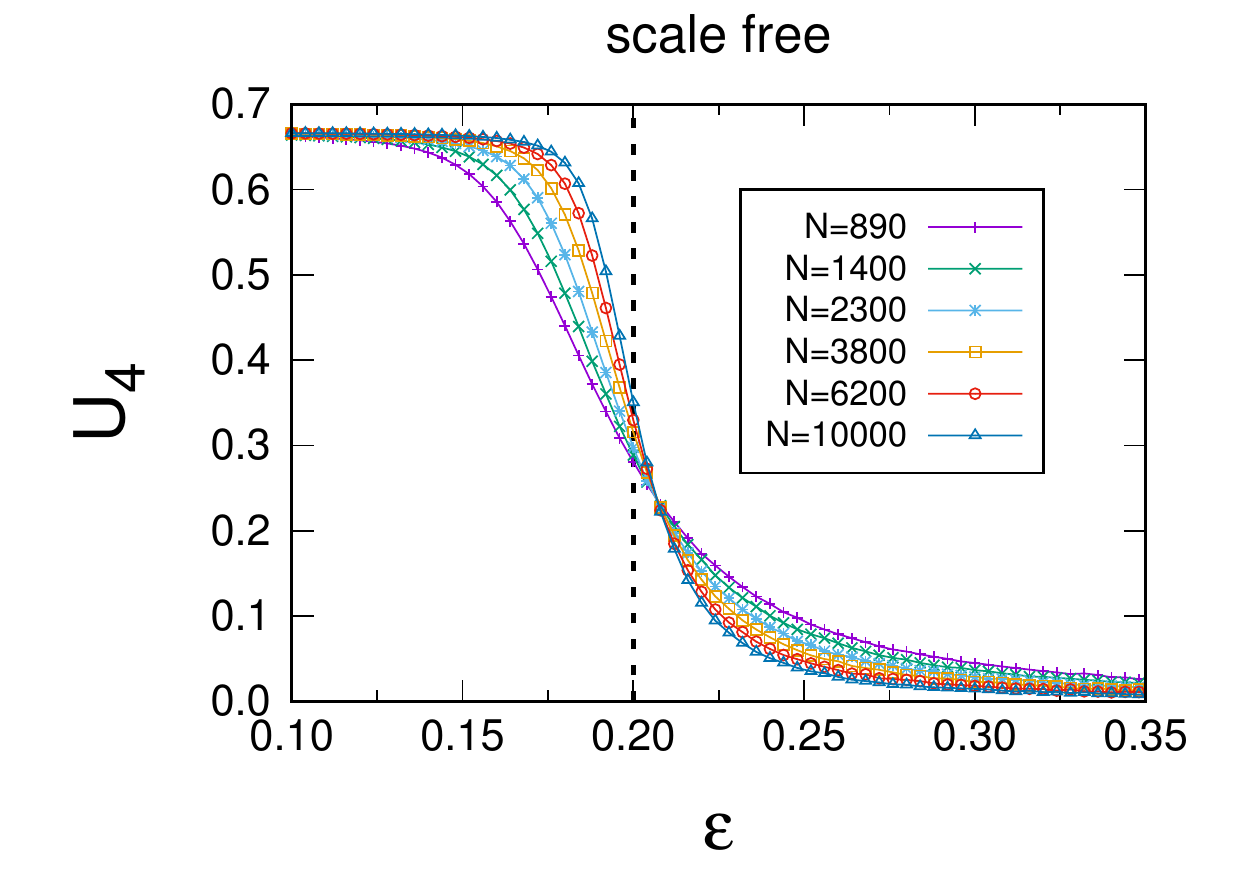}
\includegraphics[width=0.32\textwidth]{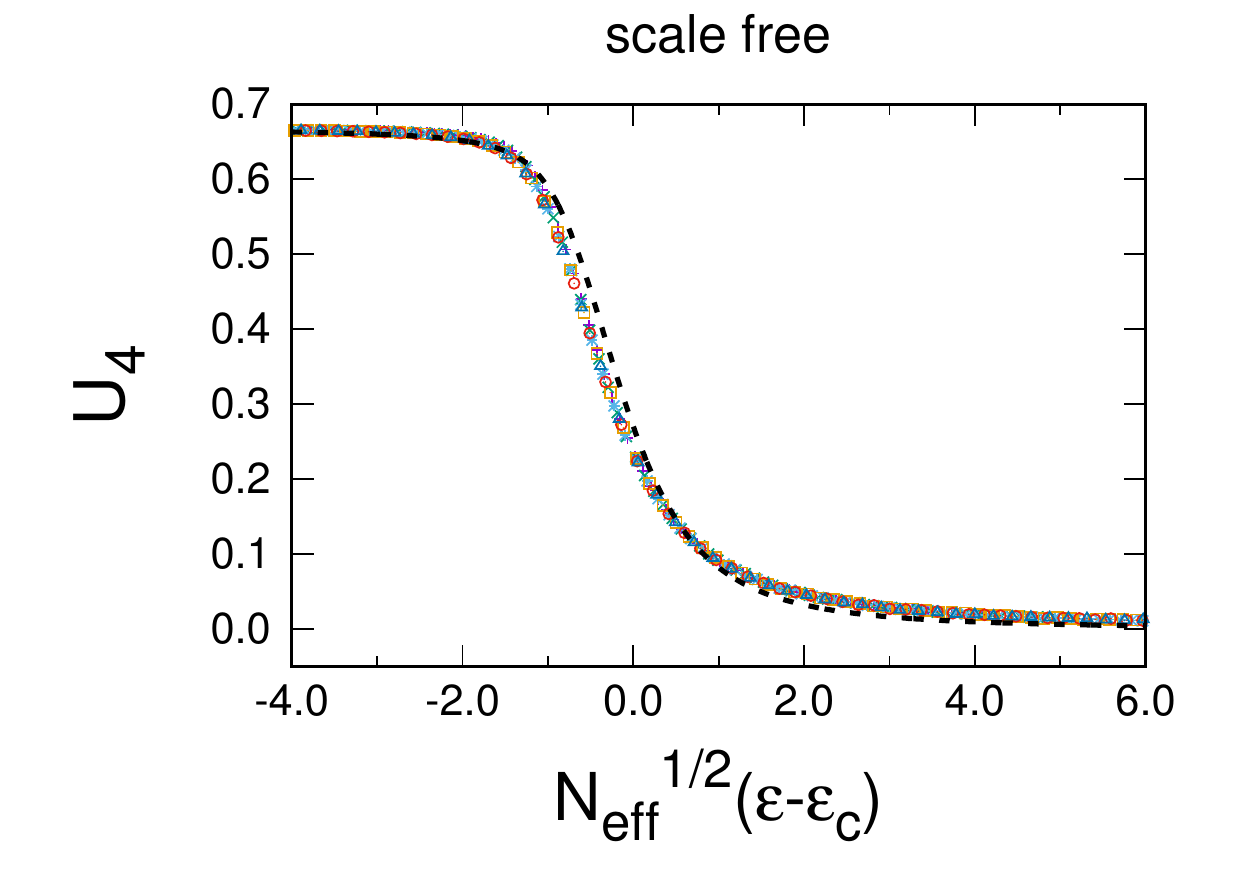}
\includegraphics[width=0.32\textwidth]{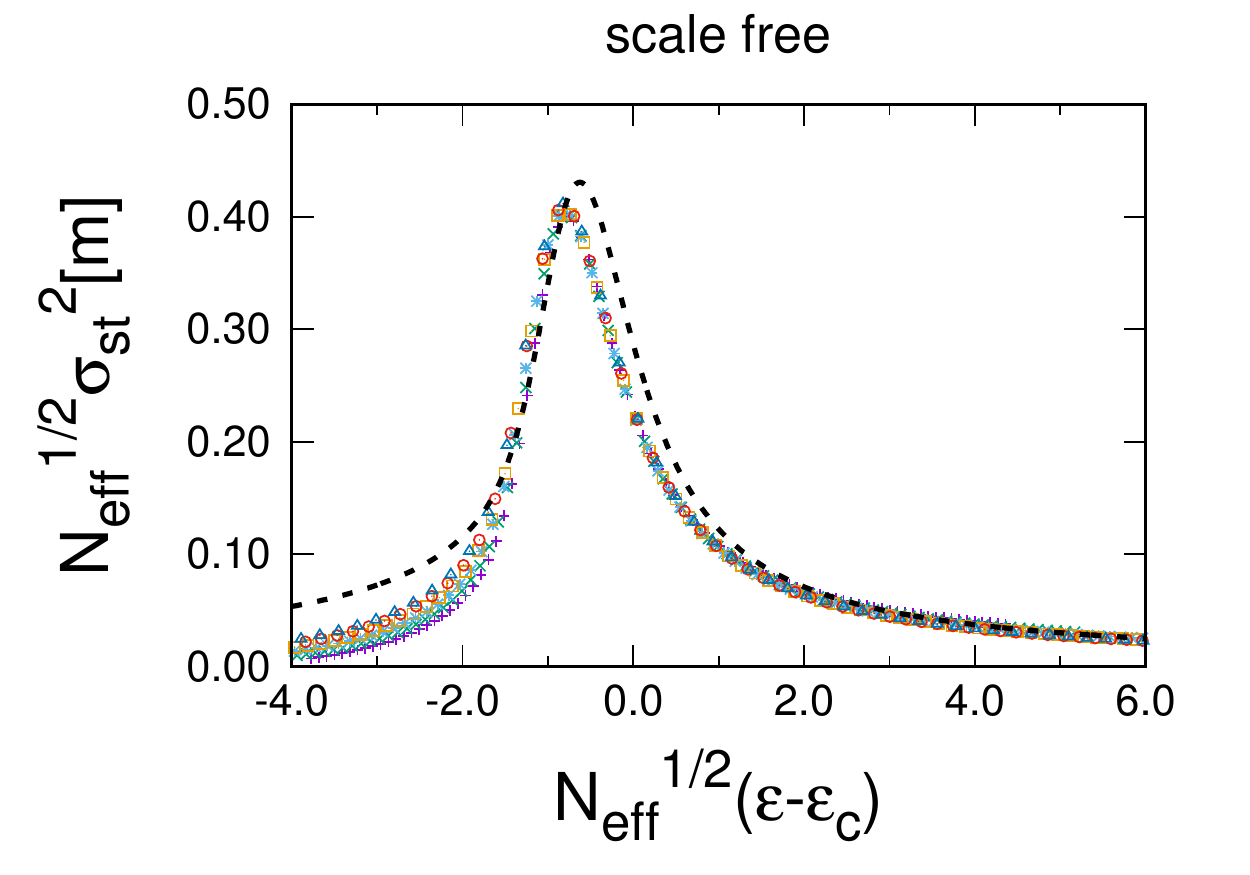}
\caption{Numerical and theoretical results for the 15-regular network and a scale free network with $P_{k \geq 6} \sim k^{-2.51}$ and $\mu \approx 14.79$, $\mu_{-1} \approx 0.11$. Dots correspond to numerical simulations of different system sizes averaged over $10^6$ Monte Carlo steps and an ensemble of $100$ networks, while the dashed black lines are the prediction of the pair-approximation which are: $\varepsilon_{c}=0.201$, $c=1.15$ and $c_4=0.12$ for the 15-regular network and $\varepsilon_{c}=0.200$, $\bar{c}=1.19$ and $\bar{c}_4=0.12$ for the scale free network.}
\label{fig:results15}
\end{figure*}

There are at least three possible sources of the quantitative discrepancy between the scaling function derived from the theory and the numerical results: (i) the scale-free network has no cut-off so that the maximum degree scales as $k_{\text{max}} \sim N^{1/(\lambda-1)}$ and there will be degree-degree correlations which are neglected in the pair-approximation; (ii) the truncation of expansion Eq.(\ref{2Fokker-Planck_equation}) may not converge rapidly with $\Delta_{k}$ for the scale free network, producing anomalous diffusion; and (iii) the binomial ansatz for the probabilities $P_{0/1}(k,q)$ fails in two different ways for low degrees $k$. First, the single event probabilities $p_0$ and $p_1$ depend also on the degree (see \citep{Pugliese}) and, second, the shape of the distribution may differ from being binomial (see \citep{Gleeson2}). In order to elucidate which one is more important we repeated the analysis for highly connected networks, which are equally affected by the error sources (i) and (ii) compared to the previous low average degree networks. The corresponding results are presented in Fig.\ref{fig:results15}. We explored a 15-regular network, and a scale free network with $P_{k \geq 6} \sim k^{-2.51}$ and average degree $\mu \approx 14.79$. As one can appreciate in the figure, the theoretical predictions are greatly improved in this case, which indicates that the third cause is the most significant one. There are still some discrepancies for the scale free network, which are caused by the other two possible error sources mentioned above.
 
\begin{figure*}[h]
\centering
\includegraphics[width=0.40\textwidth]{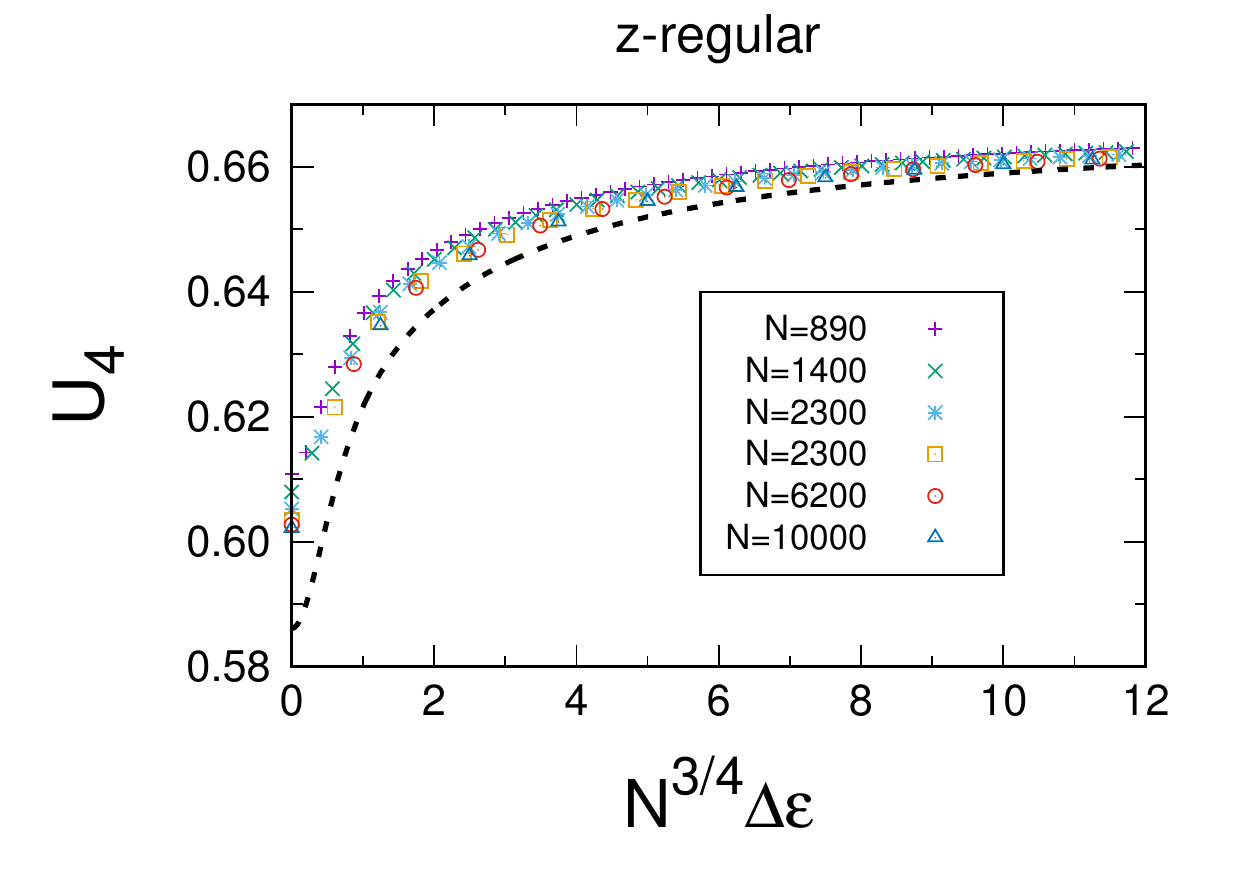}
\includegraphics[width=0.40\textwidth]{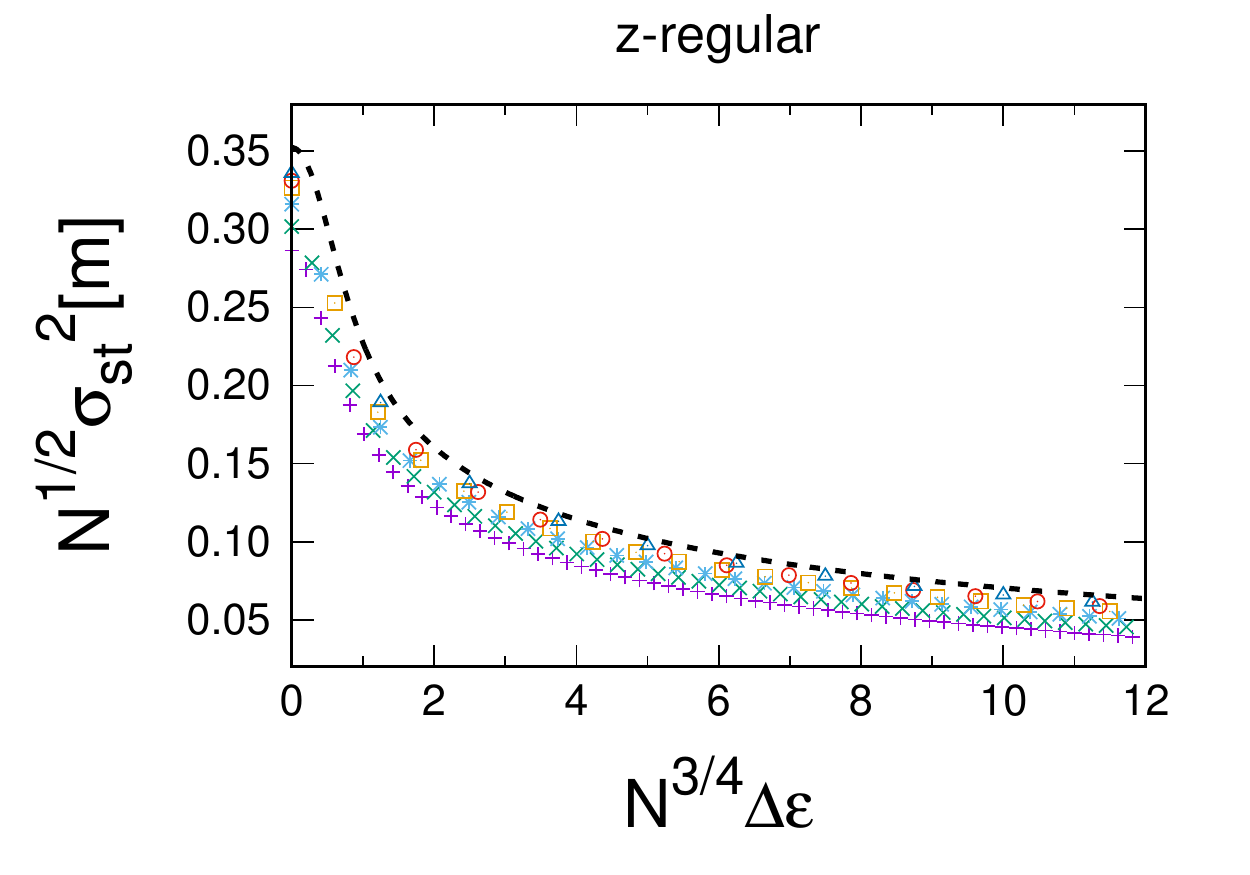}
\includegraphics[width=0.40\textwidth]{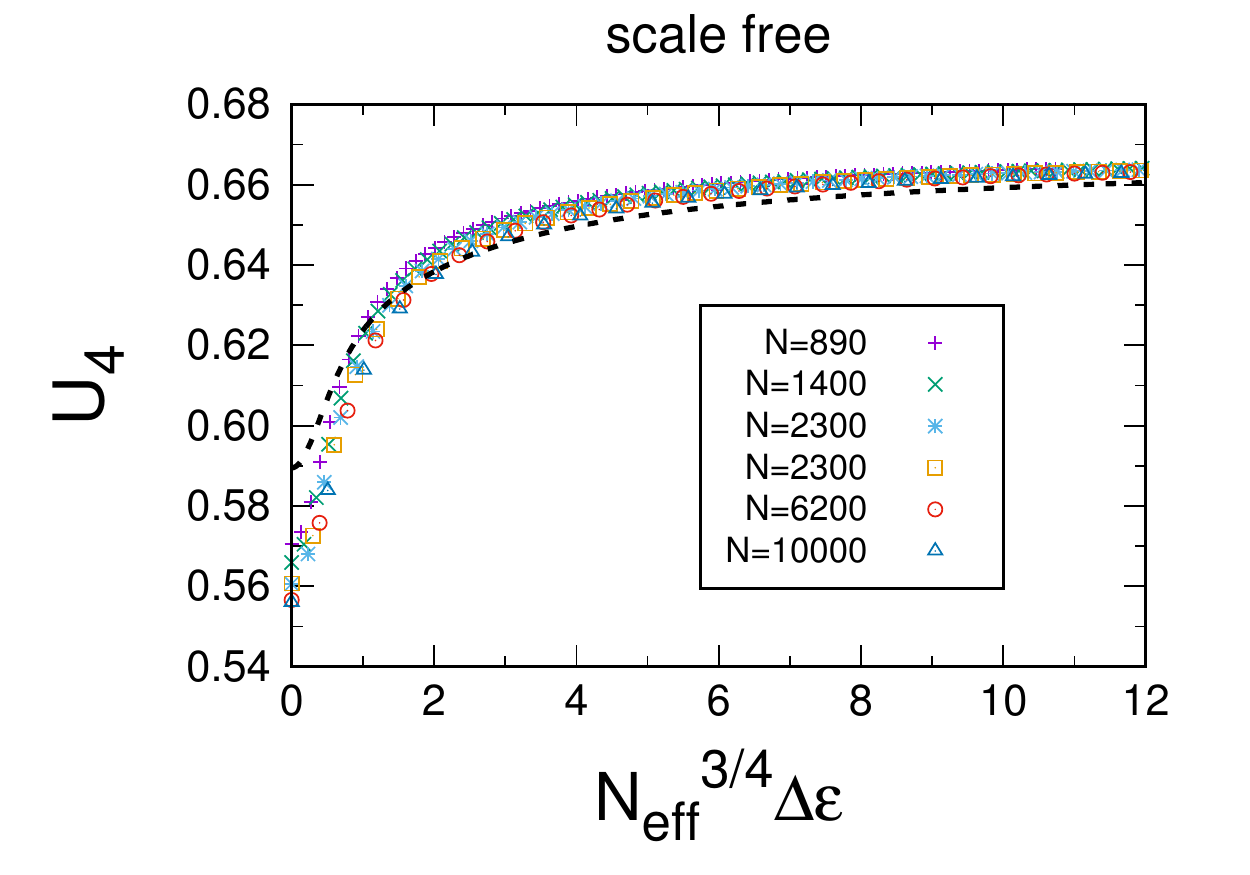}
\includegraphics[width=0.40\textwidth]{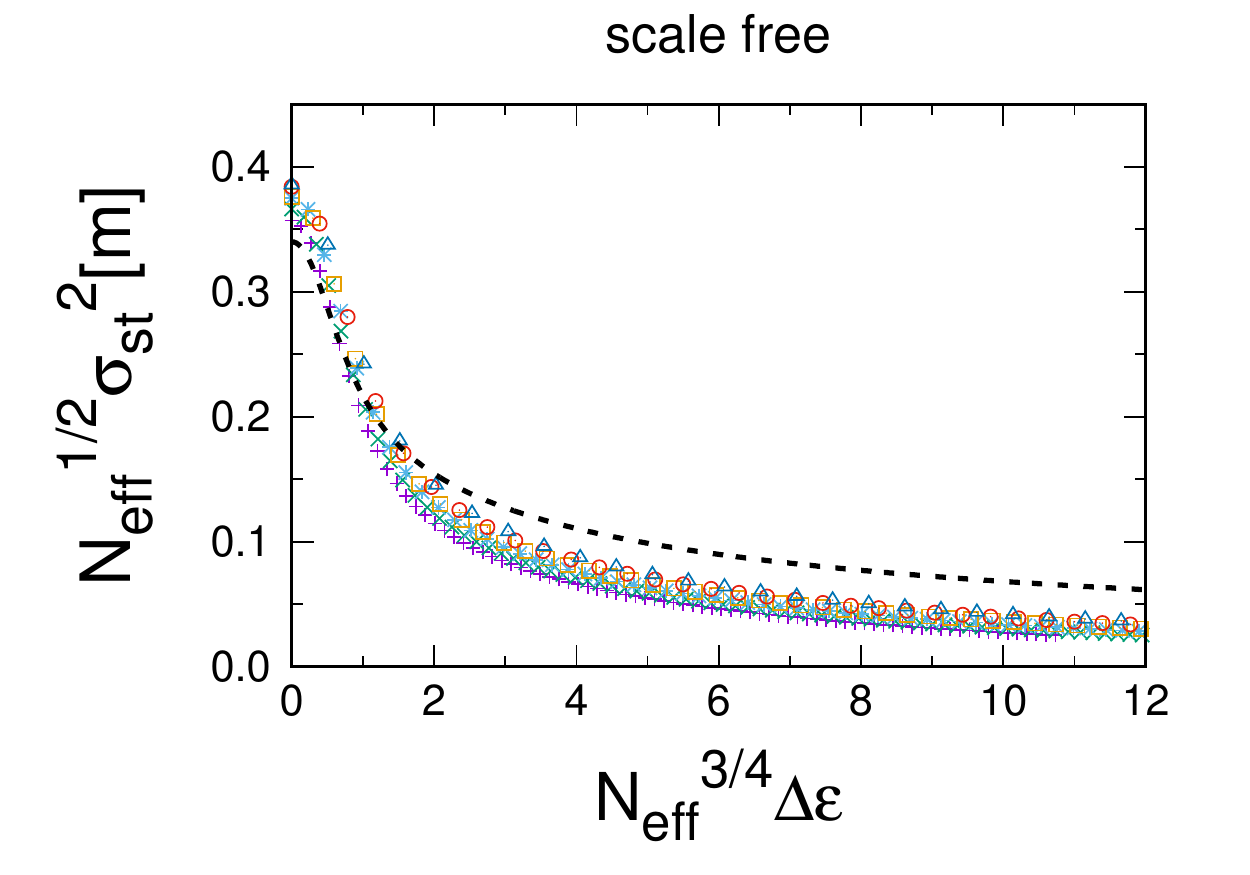}

\caption{Binder cumulant and rescaled magnetization variance as a function of the rescaled asymmetry parameter $N^{3/4}\Delta \varepsilon$ for the z-regular network, and $N_{\text{eff}}^{3/4}\Delta \varepsilon$ for the scale free network. The simulation and network specifications, together with the theoretical results are the same of those of Fig. \ref{fig:results15}. The value of $\varepsilon$ is taken to be below the critical value, $\varepsilon = \varepsilon_{c} - 1/N^{1/2}$ for the 15-regular network and $\varepsilon = \varepsilon_{c} - 1/N_{\text{eff}}^{1/2}$ for the scale free network, so that as the asymmetry increases the system crosses the catastrophe lines. The corresponding scaling functions are calculated using Eq.(\ref{phi_k}) taking into account that $x=-1$ is fixed and the variable in this case is the asymmetry $y$.}
\label{fig:results_asy}
\end{figure*}

In Fig. \ref{fig:results_asy} we additionally checked if the theoretical analysis is also valid for the asymmetric case. For the highly connected networks we take values of $\varepsilon \leq \varepsilon_{c}$ below the critical point previously calculated in Fig.\ref{fig:results15}, and vary the asymmetry parameter $\Delta \varepsilon$ checking the scaling law Eq.(\ref{moment_k2}). The results suggest that the theory is equally accurate for the asymmetric case, with a very different tendency of the moments with respect to $\Delta \varepsilon$ before and after the catastrophe.

As in the scale-free network we have $N_\text{eff}\sim N^{1-b}$, the scaling laws Eq.(\ref{moment_k2}) can be written in terms of the physical system size $N$ as
\begin{equation}
\label{moment_k2c}
\langle m^{k} \rangle_{\text{st}} = N^{-k\beta/{\bar \nu}}\tilde{\phi}_{k} \left[ N^{1/\bar\nu} (\varepsilon -\varepsilon _{c}), N^{\delta\beta/\bar\nu} \Delta\varepsilon \right],
\end{equation}
with an appropriate scaling function $\tilde{\phi}_k$ and $\beta=1/2,\,\delta=3,\,1/\bar\nu=(1-b)/2$. In the thermodynamic limit, one recovers the Landau theory mean-field exponents $\langle |m|\rangle\sim(\varepsilon_c-\varepsilon)^{1/2}$ for $\Delta\varepsilon=0$, and $\langle |m|\rangle\sim(\Delta\varepsilon)^{1/3}$ for $\varepsilon=\varepsilon_c$. For the normalized fluctuations (the equivalent of the ``magnetic susceptibility'') $\chi=N[\langle m^2 \rangle_{\text{st}}-\langle \vert m \vert \rangle_{\text{st}}^2]$ we find the scaling behavior $\chi(\varepsilon,N)=N^{\gamma/\bar\nu}\bar\chi \left[ N^{1/\bar\nu} (\varepsilon -\varepsilon _{c}), N^{\delta\beta/\bar\nu} \Delta\varepsilon \right]$ with a scaling function $\bar\chi$ and $\gamma/\bar\nu=1-2\beta/\bar\nu$, and $\gamma=1$. Again, in the thermodynamic limit we find the mean-field result $\chi\sim |\varepsilon-\varepsilon_c|^{-1}$. Therefore, although the critical exponents of the transition are those given by the mean-field theory, the analysis in terms of the finite system size $N$ yields a network-dependent exponent $1/\bar\nu=(1-b)/2$. We have computed $b=0.36(1)$ for the scale free network with tail exponent $\lambda=2.43$, yielding $1/\bar\nu=0.160(5)$ and $\gamma/\bar\nu=1-2\beta/\bar\nu=0.680(5)$ in perfect agreement with the numerical fits $\beta/\bar\nu=0.157(6)$ and $\gamma/\bar\nu=0.691(9)$ reported in~\citep{Jedrzejewski}.

\section{Summary and Conclusions}\label{conclusions}
In this paper we have studied the impact of a non-linear herding mechanism in the noisy voter model, with the consideration of a parameter $\alpha$, the ``volatility'' in the context of the Abrams-Strogatz model~\citep{Strogatz}, that measures the resistance of individuals to copy the state of a neighbor in the opposite state or the effect of complex, collective, interactions. $\alpha$ is taken as a real parameter that can take any non-negative value. The solution in the all-to-all scenario, where all voters are neighbors of each other, shows that the well studied bimodal-unimodal finite-size transition of the noisy voter model turns into a classical second-order transition, with a critical point $\varepsilon_{c} = 2^{-\alpha}(\alpha-1) \neq 0$ for $\alpha>1$, no transition for $\alpha<1$, and a finite-size transition $\varepsilon_{c}(N)=1/N$ for the traditional linear case $\alpha=1$. In the strong non-linear regime $\alpha>5$, the transition $\varepsilon_{c}$ separates bimodal from trimodal states, with the appearance of a new first order transition line $\varepsilon_{t} > \varepsilon_{c}$ that separates trimodal from unimodal states. This implies that this non-linear mechanism is able to generate tristability and the point where the tristability parameter region begins is a tricritical point $\alpha=5$, $\varepsilon = \varepsilon_{c} = \varepsilon_{t}$, in accordance with the results of~\citep{Nyczka}, valid for integer $\alpha$. The asymmetry in the rates of the model $\Delta\varepsilon \neq 0$ is capable of destroying bimodality and trimodality at what is known as cusp and butterfly catastrophes respectively.

The existence of a {\slshape bona-fide} transition that remains in the thermodynamic limit offers a convenient solution to the $N$-dependence of the transition without the need to resort to a $N$-dependent rescaling of the model parameters~\citep{Alfarano2}, also known as non-extensive formulation. Although this formulation partially solves this problem, it makes the model structurally unstable in such a way that simple perturbations such as the inclusion of agents that do not change state~\citep{Nagi,Kononovicius,Kononovicius2}, an external signal of information~\citep{Carro1}, or a network structure, can drastically change the properties to be Gaussian~\citep{Alfarano3}. Similar conclusions were already reported in the context of the voter model~\citep{Castello1} (without spontaneous switching). The addition of a third state or a small perturbation of the functional form of the rates of the voter model drastically changes the dynamics and the ordering mechanism. These structural changes of the voter model are specially relevant in the context of agent-based modeling of language competition~\citep{Vazquez1}. In this context, the inclusion of a third state represents the role of bilingual speakers in the evolution of the number of speakers of a particular language.

The $N$-dependence of the non-linear model in the presence of noise is dramatically different from the linear case. The aforementioned non-extensive formulation of the linear model leads to $N$-independent statistics of the magnetization $\langle m^{k} \rangle \sim N^{0}$. For the non-linear case, however, we have shown that rescaling the parameters leads to $\langle m^{k} \rangle \sim N^{-k/4}$ for $1<\alpha<5$ at the critical point $\varepsilon_c$ and $\langle m^{k} \rangle \sim N^{-k/6}$ at the tricritical point for $\alpha=5$, i.e. vanishing moments in the thermodynamic limit $N \rightarrow \infty$. The critical region where these scaling laws are valid, however, is wider for the non-linear case making them more robust against perturbations.

We have also checked the role of the network of interactions in the presented results of the model, making use of the pair approximation. It would indeed be possible to develop higher accuracy theoretical methods at the cost of simplicity and analytical tractability~\citep{Pugliese, Gleeson1, Gleeson2}. In general, we conclude that the critical point $\varepsilon_{c}$ is lowered compared to the all-to-all solution, depending on the mean degree $\mu$ and the first negative moments of the degree distribution $\mu_{-1}$,..., $\mu_{-\alpha+1}$ (for $\alpha$ integer). The $N$-dependent scaling laws are also changed for networks whose second moment of the degree distribution $\mu_2$ changes dramatically with system size, e.g. scale-free networks $\mu_2 \sim N^{b}$ with $0<b<1$. In this case, we have shown that one can define an effective system size $N_{\text{eff}} = N\mu^2/\mu_2 \sim N^{1-b}$ for which the all-to-all scaling laws become valid. In the limit of highly heterogeneous networks, $\lambda \rightarrow 2$ and $b \rightarrow 0$, statistics may be $N$-independent or with a very weak dependence on system size.

\section*{Acknowledgements}
We acknowledge financial support from Agencia Estatal de Investigaci\'on (AEI, Spain) and Fondo Europeo de Desarrollo Regional under Project ESoTECoS Grant No. FIS2015-63628-C2-2-R (AEI/FEDER,UE). A. F. P. acknowledges support by the FPU program of MECD (Spain).


\clearpage

\section*{References}
\begin{thebibliography}{99}

\bibitem{Clifford}
P. Clifford and A. Sudbury. A model for spatial conflict. {\em Biometrika} {\bf 60}, 581 (1973).

\bibitem{Holley}
R. A. Holley and T. M Liggett. Theorems for Weakly Interacting Infinite Systems and the Voter Model. {\em Ann. Probab.} {\bf 3},
643 (1975).

\bibitem{Hammal} O. Al Hammal, H. Chat\'e, I. Dornic, and M. A. Mu\~noz.
Langevin Description of Critical Phenomena with Two Symmetric Absorbing States.
{\em Phys. Rev. Lett.} {\bf 94}, 230601 (2005).

\bibitem{clopez}
F. Vazquez and C. L\'opez.
\newblock {Systems with two symmetric absorbing states: Relating the microscopic dynamics
with the macroscopic behavior}.
\newblock {\em Physical Review E} \textbf{78}, 061127 (2008).

\bibitem{Castellano}
C. Castellano, M. A. Mu\~noz, and R. Pastor-Satorras.
\newblock {Nonlinear \emph{q}-voter model}.
\newblock {\em Physical Review E} \textbf{80}, 041129 (2009).

\bibitem{Granovsky}
B. L. Granovsky and N. Madras.
\newblock {The noisy voter model}.
\newblock {\em Stochastic Processes and their Applications} \textbf{55}, 23-43 (1995).

\bibitem{Kirman}
A. Kirman.
\newblock {Ants, Rationality, and Recruitment}.
\newblock {\em The Quarterly Journal of Economics} \textbf{108}, 137-156 (1993).

\bibitem{Moran}
P. A. P. Moran.
Random processes in genetics. {\em Mathematical Proceedings of the Cambridge Philosophical Society}, Vol. 54. No. 1. Cambridge University Press (1958).

\bibitem{Lebowitz}
J. L. Lebowitz and H. Saleu. Percolation in strongly correlated systems. {\em Physica A} {\bf 138}, 194 (1986).

\bibitem{Fichthorn}
K. Fichthorn, E. Gulari, and R. Ziff. Noise-induced bistability in a Monte Carlo surface-reaction model. {\em Phys. Rev. Lett.} {\bf 63}, 1527 (1989).

\bibitem{Considine}
D. Considine, S. Redner, and H. Takayasu. Comment on ``Noise-induced bistability in a Monte Carlo surface-reaction model''. {\em Phys. Rev. Lett.} {\bf 63}, 2857 (1989).

\bibitem{Diakonova}
M. Diakonova, V. M. Eguiluz, and M. San Miguel.
\newblock{Noise in Coevolving Networks}.
\newblock{\em Physical Review E} \textbf{92}, 032803 (2015).

\bibitem{Alfarano1}
S. Alfarano, T. Lux, and F. Wagner.
\newblock {Estimation of Agent-Based Models: The Case of an Asymmetric Herding Model}.
\newblock {\em Computational Economics} \textbf{26}, 19-49 (2005).

\bibitem{Alfarano2}
S. Alfarano, T. Lux, and F. Wagner.
\newblock {Time variation of higher moments in a financial market with heterogeneous agents: An analytical approach}.
\newblock {\em Journal of Economic Dynamics} \& {\em Control} \textbf{32}, 101-136 (2008).

\bibitem{Schweitzer}
F. Schweitzer and L. Behera.
\newblock {Nonlinear voter models: the transition from invasion to coexistence}.
\newblock {\em European Physical Journal B} \textbf{67}, 301-318 (2009).

\bibitem{Strogatz}
D. M. Abrams and S. H. Strogatz.
\newblock {Modelling the dynamics of language death}.
\newblock {\em Nature} \textbf{424}, 900 (2003).

\bibitem{Vazquez1}
F. Vazquez, X. Castell\'o, and M. San Miguel.
\newblock {Agent based models of language competition: macroscopic descriptions and order-disorder transitions}.
\newblock {\em Journal of Statistical Mechanics: Theory and Experiment} \textbf{2010}(04), P04007 (2010).

\bibitem{Nowak}
A. Nowak, J. Szamrej, and B. Latan\'e.
\newblock{From private attitude to public opinion: A dynamic theory of social impact}.
\newblock{\em Psychological Review} \textbf{97}, 362 (1990).

\bibitem{Nettle}
D. Nettle.
\newblock{Using social impact theory to simulate language change}.
\newblock{\em Lingua} \textbf{108}, 95 (1999).

\bibitem{Min}
B. Min and M. San Miguel.
\newblock{Fragmentation transitions in a coevolving nonlinear voter model}.
\newblock{\em Scientific Reports} \textbf{7}, 12864 (2017).

\bibitem{Nyczka}
P. Nyczka, K. Sznajd-Weron, and J. Cis{\l}o.
\newblock{Phase transitions in the $q$-voter model with two types of stochastic driving}.
\newblock{Physical Review E} \textbf{86}, 011105 (2012).

\bibitem{Nyczka-2}
P. Nyczka and K. Sznajd-Weron.
Anticonformity or Independence?-Insights from Statistical Physics
{\em J. Stat. Phys.} {\bf 151}, 174 (2013).

\bibitem{Jedrzejewski}
A. J\polhk{e}drzejewski.
\newblock{Pair approximation for the $q$-voter model with independence on complex networks}.
\newblock{Physical Review E} \textbf{95}, 012307 (2017).

\bibitem{Moretti} P. Moretti, S. Liu, C. Castellano, and R. Pastor-Satorras.
Mean-Field Analysis of the q-Voter Model on Networks.
{\em J. Stat. Phys.} {\bf 151}, 113 (2013).

\bibitem{Lambiotte} R. Lambiotte and S. Redner.
Dynamics of non-conservative voters.
EPL {\bf 82}, 18007  (2008).

\bibitem{Molofsky} J. Molofsky, R. Durrett, J. Dushoff, D. Griffeath, and S. Levine.
Local Frequency Dependence and Global Coexistence.
{\em Theoretical Population Biology} {\bf 55}, 270 (1999)

\bibitem{Huang}
K. Huang.
\newblock {\em Statistical Mechanics}.
\newblock John Wiley \& Sons. New York (1963).

\bibitem{Escaff}
D. Escaff, R. Toral, C. Van den Broeck, and K. Lindenberg.
\newblock{A continuous-time persistent random walk model for flocking}.
\newblock{arXiv:1803.02114}.

\bibitem{Konstantin}
M. A. Serrano, K. Klemm, F. Vazquez, V. M. Eguiluz, and M. San Miguel.
Conservation laws for voter-like models on random directed networks.
{\em Journal of Statistical Mechanics: Theory and Experiment} {\bf 2009}(10), P10024 (2009).

\bibitem{Toral-Colet:2014}
R. Toral and P. Colet.
\newblock {\em {Stochastic Numerical Methods: An Introduction for Students and Scientists}}.
\newblock Wiley, 2014.

\bibitem{VanKampen:2007}
N. G. van Kampen.
\newblock {\em Stochastic Processes in Physics and Chemistry}.
\newblock {North-Holland Physics Publishing}: Amsterdam, 2007.

\bibitem{Cata1}
E. C. Zeeman.
\newblock {Catastrophe Theory}.
\newblock {\em Scientific American} \textbf{234}, 65-83 (1976).

\bibitem{Cata2}
R. Gilmore.
\newblock {\em Catastrophe Theory for Scientists and Engineers}. New York: Wiley (1981).

\bibitem{Alfarano3}
S. Alfarano and M. Milakovi\'c.
\newblock {Network structure and $N$-dependence in agent-based herding models}.
\newblock {\em Journal of Economic Dynamics} \& {\em Control} \textbf{33}, 78-92 (2009).

\bibitem{Gleeson1}
J. P. Gleeson.
\newblock {High-Accuracy Approximation of Binary-State Dynamics on Networks}.
\newblock {\em Phys. Rev. Lett.} \textbf{107}, 068701 (2011).

\bibitem{Gleeson2}
J. P. Gleeson.
\newblock {Binary-State Dynamics on Complex Networks: Pair Approximation and Beyond}.
\newblock {\em Physical Review X} \textbf{3}, 021004 (2013).

\bibitem{Carro2}
A. Carro, R. Toral, and M. San Miguel.
\newblock{The noisy voter model on complex networks}.
\newblock{Scientific Reports} \textbf{6}, 24775 (2016).

\bibitem{Peralta}
A. F. Peralta, R. Toral, A. Carro, and M. San Miguel.
\newblock{Stochastic pair approximation treatment of the noisy voter model}.
\newblock{Preprint 2018}.

\bibitem{Redner}
V. Sood, T. Antal, and S. Redner.
\newblock {Voter models on heterogeneous networks}.
\newblock {\em Physical Review E} \textbf{77}, 041121 (2008).

\bibitem{Pugliese}
E. Pugliese and C. Castellano.
\newblock {Heterogeneous pair approximation for voter models on networks}.
\newblock {\em A Letters Journal Exploring the Frontiers of Physics} \textbf{88}, 58004 (2009).

\bibitem{Vazquez2}
F. Vazquez and V. M. Eguiluz.
\newblock {Analytical solution of the voter model on uncorrelated networks}.
\newblock {\em New Journal of Physics} \textbf{10}, 063011 (2008).

\bibitem{Romualdo}
M. Catanzaro, M. Bogu\~n\'a, and R. Pastor-Satorras.
\newblock {Generation of uncorrelated random scale-free networks}.
\newblock {\em Physical Review E} \textbf{71}, 027103 (2005).

\bibitem{Binder}
K. Binder.
\newblock{Finite-size Scaling Analysis of Ising Model Block Distribution Functions}.
\newblock{ \em Z. Phys. B - Condensed Matter} \textbf{43}, 119 (1981).

\bibitem{Carro1}
A. Carro, R. Toral, and M. San Miguel.
\newblock {Markets, herding and response to external information}.
\newblock{PLoS ONE} \textbf{10(7)}:e0133287 (2015).

\bibitem{Nagi}
N. Khalil, M. San Miguel, and R. Toral.
\newblock{Zealots in the mean-field noisy voter model}.
\newblock{\em Physical Review} E \textbf{97}, 0123101 (2018).

\bibitem{Kononovicius}
A. Kononovicius and V. Gontis.
Control of the socio-economic systems using herding interactions.
{\em Physica A} {\bf  405}, 80 (2014).

\bibitem{Kononovicius2}
A. Kononovicius and V. Gontis.
Herding interactions as an opportunity to prevent extreme events in financial markets.
{\em Eur. Phys. J. B} {\bf 88}, 189  (2015).

\bibitem{Castello1}
X. Castell\'o, V. M. Eguiluz, and M. San Miguel.
\newblock {Ordering dynamics with two non-excluding options: bilingualism in language competition}.
\newblock {\em New Journal of Physics} \textbf{8}, 308 (2006).

\bibitem{f1}The simulations proceed by selecting randomly one of the $N$ nodes, say $i$. If $n_i=0$ the transition $n_i\to1$ occurs with a probability $\pi_i^+/C$ and if $n_i=1$ the transition $n_i\to0$ occurs with probability $\pi_i^-/C$ with $C=\max(a_0,a_1)+h$. $N$ node selections constitute one Monte Carlo step~\citep{Toral-Colet:2014}.

\bibitem{f2}There is also a recurrence relation for the master equation in the stationary state: $P_{\rm st}(n)=\dfrac{\pi^+(n-1)}{\pi^-(n)}P_\text{st}(n-1)$, but for a general value of $\alpha$ it does not seem to be possible to reduce this expression to an analytically tractable function.

\bibitem{f3}There is a misprint in that reference~\citep{Vazquez1} and the factor $(\alpha-9)$ is missing in their formula 10.

\bibitem{f4}In some interpretations of the model, the ordered phase corresponds to social consensus where a large fraction of the population holds the same option for the variable $n_i$.

\bibitem{directed}In the case of directed networks, there is an equation for $\rho_{01}$ and another one for $\rho_{10}$, which are formally similar to Eq.(\ref{rho_evolution1}). The difference is that the changes of these quantities are $\Delta \rho^{01}_{k,q ; k_{\text{in}}, q_{\text{in}}} = 2 (k_{\text{in}}-q_{\text{in}}-q)/\mu N$ and $\Delta \rho^{10}_{k,q ; k_{\text{in}}, q_{\text{in}}} = 2 (k-q-q_{\text{in}})/\mu N$, which depend on the in-degree $k_{i}^{\text{in}} = \sum_{j} A_{ji}$ which is different from the out-degree $k_{i} = \sum_{j} A_{ij}$. The single event probabilities of $P_{0/1}(k,q)$ are in this case $p_{0}=\rho_{01}$ and $p_{1}=1-\rho_{10}$, and we have to define the binomial $P^{\text{in}}_{0/1}(k_{\text{in}},q_{\text{in}})$ with single event probabilities $p^{\text{in}}_{0}=\rho_{10}$ and $p^{\text{in}}_{1}=1-\rho_{01}$. The corresponding Eq.(\ref{rho_evolution1}) will need to be averaged additionally by the in-degree distribution $P^{\text{in}}_{k_{\text{in}}}$ (or by $P^{\text{in}}_{k_{\text{in}} \vert k}$ if there are $k_{\text{in}}-k$ correlations) and $P^{\text{in}}_{0/1}(k_{\text{in}},q_{\text{in}})$. Obviously this treatment is valid only if reciprocal links are negligible, otherwise the method is significantly more involved.

\end{thebibliography}
\end{document}